\documentclass[aps,prl,reprint, superscriptaddress]{revtex4-1}

\usepackage{amsmath,amssymb, amsthm,amstext}
\usepackage{natbib}
\usepackage{graphicx}
\usepackage{color}
\usepackage{array, enumerate}
\usepackage{bm}
\usepackage{multirow}
\usepackage[breaklinks,colorlinks,citecolor=blue]{hyperref}
\usepackage{braket}
\usepackage{txfonts}

\def\be{\begin{equation}}
	\def\ee{\end{equation}}

\begin{document}
	\title{Dynamical Instability of Self-Gravitating Membranes }
	\author{Huan Yang}
	\email{hyang@perimeterinstitute.ca}
	\affiliation{Perimeter Institute for Theoretical Physics, Ontario, N2L 2Y5, Canada}
	\affiliation{University of Guelph, Guelph, Ontario N1G 2W1, Canada}
	\author{B\'eatrice Bonga}
	\email{bbonga@science.ru.nl}
	\affiliation{Institute for Mathematics, Astrophysics and Particle Physics, Radboud University, 6525 AJ Nijmegen, The Netherlands}
	\author{Zhen Pan}
	\email{zpan@perimeterinstitute.ca}
	\affiliation{Perimeter Institute for Theoretical Physics, Ontario, N2L 2Y5, Canada}

	\begin{abstract}
		We show that a generic relativistic membrane with in-plane pressure  and surface density having the same sign is unstable with respect to a series of warping mode instabilities with high wave numbers.
		We also examine the criteria of instability for commonly studied exotic compact objects with membranes, such as gravastars, AdS bubbles and  thin-shell wormholes.
		For example, a gravastar which satisfies the weak energy condition turns out to be dynamically unstable. A thin-layer black hole mimicker is stable only if it has positive pressure and negative surface density (such as a wormhole), or vice versa.
	\end{abstract}
	\maketitle

\noindent{\bf Introduction.} The detection of binary black hole (BH) mergers with ground-based gravitational-wave (GW) detectors \cite{LIGOScientific:2016aoc,LIGOScientific:2018mvr,LIGOScientific:2020ibl,LIGOScientific:2021djp}, images of the supermassive BHs M87 $\&$ Sgr A* with radio interferometry \cite{M87PaperI,EventHorizonTelescope:2022xnr}, and the observation of S stars orbiting a small dark region in the galactic center \cite{schodel2002star}, all point to the existence of BHs, of which a description was first obtained by  Karl Schwarzschild more than one hundred years ago using General Relativity. The study of BHs is not limited to astrophysics and General Relativity, but also plays a role in other major areas of physics, such as quantum fields and strings, condensed matter physics and quantum information. Because of its unparalleled conceptual and observational importance, it is paramount to test the more refined features of BHs against all viable alternatives allowed by the laws of nature. Any signal, e.g. ringdown quasinormal modes \cite{Berti:2016lat,Yang:2017zxs},  that favors a BH mimicker over BHs themselves would represent a fundamental breakthrough/revolution in physics. In the coming decades the third-generation ground-based GW detectors \cite{Punturo:2010zz,reitze2019cosmic}, the space-borne GW detectors \cite{Audley:2017drz,TianQin:2015yph} and the next-generation Event Horizon Telescope, will likely improve the precision of such tests by many orders.

Horizonless compact objects are important candidates for BH mimickers \cite{Cardoso:2019rvt}. One class of them, such as boson stars, has smooth distributions of matter/fields that are convenient for stability analysis and numerical simulations. However, it appears difficult to construct stable configurations of these compact stars that approach the compactness of BHs. For example, a fluid star with causal equation of state can achieve maximum compactness at around $M/R \le 0.355$ \cite{haensel1989submillisecond,koranda1997upper}, with $M$ its mass and $R$ its radius as measured from the surface area. The bound for boson stars is around $0.44$ \cite{Kesden:2004qx}.
There are proposals for constructing compact stars with anisotropic stress \cite{Bowers:1974tgi,Bayin:1982vw,Dev:2000gt,Mak:2001eb,Herrera:2004xc,Yagi:2015hda,Raposo:2018rjn} to increase the maximum compactness, but they often feature problems such as superluminal sound speed, violation of energy conditions  and lack of stability analysis. A recent study showed that the bound can be improved to $\sim 0.376$ by including various prescriptions of elastic stress \cite{Alho:2022bki}.

Another class of compact objects often include a (or multiple) membrane(s) that separates spacetime regions, such as gravastars \cite{Mazur2001} and thin-shell wormholes \cite{Visser1989,Visser1989b}. These are interesting because this type of construction allows the transition to the exterior spacetime, which is the same as the BH spacetime, to be arbitrarily close to the horizon of a corresponding BH. Therefore, these models can have compactness arbitrarily close to that of BHs. In addition, 
membranes are often invoked if there is interesting physics happening near a certain surface, such as proposals considering hard structures near BH horizons motivated by firewalls or $2-2$ holes \cite{Holdom:2016nek,Kaplan:2018dqx}.
Moreover, compact objects with membranes are expected to have distinct strong-gravity dynamic behavior from more uniform compact objects. For this latter reason, two-dimensional domain walls have been extensively studied in cosmology.

Here, we present a perturbation study of self-gravitating membranes with nontrivial energy and stress. We find that if the signs of the in-plane pressure and  surface density in the membrane are the same, there is a generic warping instability for modes with sufficiently high wave numbers.
We apply these results to commonly studied compact objects, and find that a significant portion of the parameter space of gravastars --- which are usually modeled by a de Sitter interior and Schwarzschild exterior with a spherical shell of matter at the boundary  --- and AdS bubbles (with anti-de Sitter interiors) are dynamically unstable.
Static thin-shell wormholes always have positive pressure and negative surface density, so that they are free from these instabilities.
Therefore, requiring membranes to have negative pressure (for positive surface density) and positive pressure (for negative density) becomes a powerful qualifier for the stability of compact objects.
Throughout this work, we adopt geometric units $(c= G =1)$.

\begin{figure}
				\includegraphics[width=0.48\textwidth]{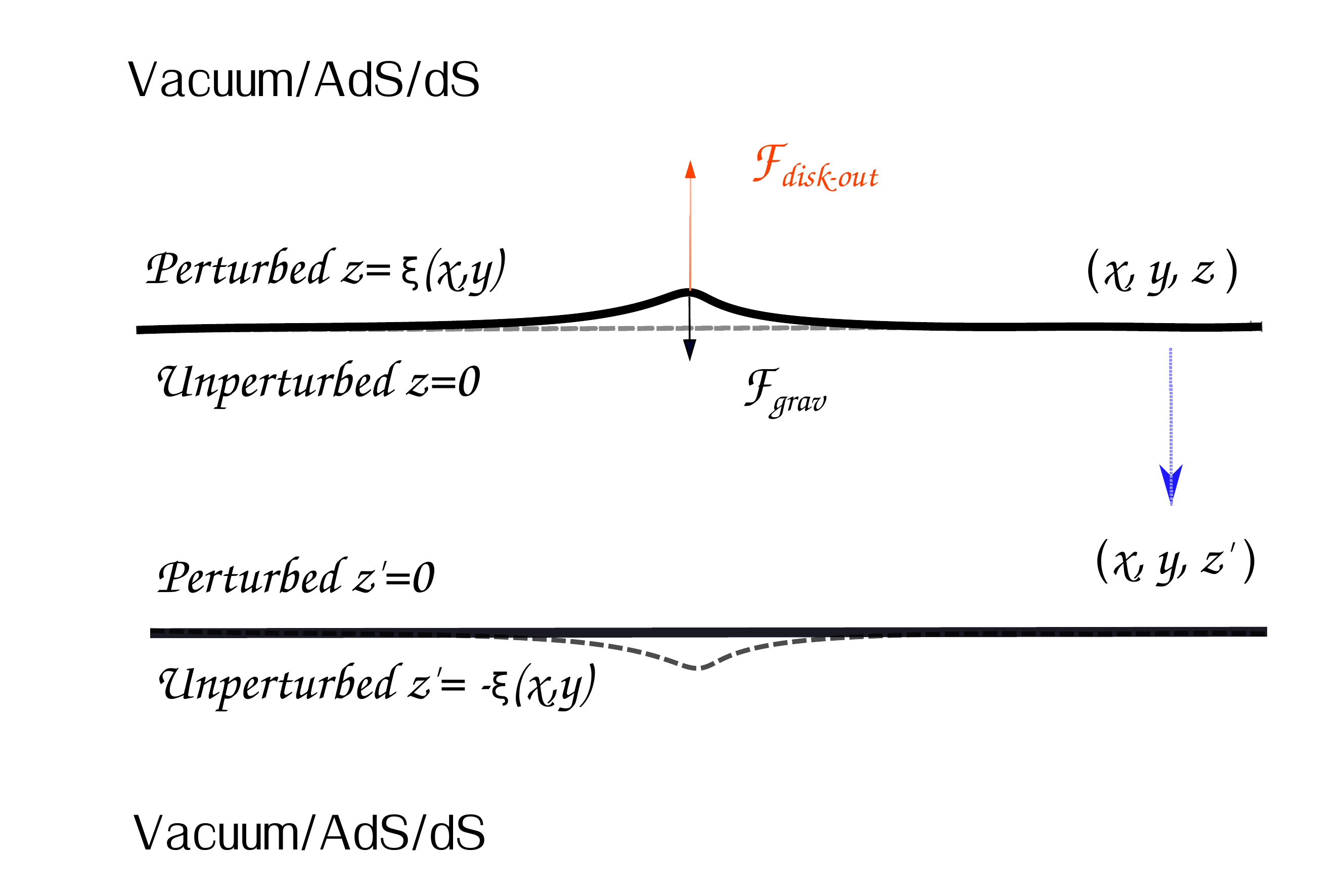}

			\caption{A self-gravitating membrane separating two spacetime regions, which are vacuum solutions to Einstein's equations with possibly a non-zero cosmological constant. The positive pressure generally produces an anti-spring force out of the plane and the gravitational pull may act as a spring force trying to bring back the displacement to its equilibrium position. For the analysis presented in both the
			Newtonian and relativistic regime, we use a coordinate transformation to map the membrane to the equatorial plane of the new coordinates to facilitate the derivation of the equation of motion.}
			\label{fig:illu}
		\end{figure}

\vspace{0.2cm}

\noindent{\bf Membrane instability.} Let us consider a self-gravitating membrane with intra-surface pressure surrounded by vacuum. If the pressure is positive, any local vertical displacement results in an ``anti-restoring" force that pushes the mass element away from equilibrium, see Fig.\ref{fig:illu}. On the other hand,  the gravitational attraction from surrounding mass elements tends to bring it back to equilibrium. We shall show that the anti-spring force always wins in the eikonal limit, leading to a series of instabilities with high wave number. To illustrate the basic picture, we present the analysis in the Newtonian regime first before proceeding to the relativistic case.

Consider a membrane placed in the ($x-y$) plane, with surface density $\sigma$ and surface pressure $P$. The displacement field $\boldsymbol{\xi}(x,y)$ can be decomposed as
\begin{align}
	\boldsymbol{\xi} = \xi_x \hat{e}_x +\xi_y \hat{e}_y +\xi_z \hat{e}_z\,.
\end{align}
We use $\delta$ to denote Eulerian perturbations and $\Delta$ to denote Lagrangian perturbations. For example, the Eulerian density fluctuation is given by $\delta \sigma =-\nabla_{\parallel} \cdot( \sigma \boldsymbol{\xi})$, where $\nabla_{\parallel}$ here operates on the two horizontal directions, and the Lagrangian density perturbation is given by $\Delta \sigma = \delta \sigma +\boldsymbol{\xi} \cdot \nabla_{\parallel} \sigma$. For the purpose of this analysis, we only need to consider the case with $\xi_z$ nonzero, in which case the local area change of mass elements is second order in $\xi$, i.e. $\delta \sigma = \Delta \sigma =0$ at linear order.

The equation of motion for \emph{three-dimensional} fluid elements, in terms of Lagrangian variables, can be written as \cite{gregorian2015nonradial}
\begin{align}\label{eq:3d}
	\rho_0 \left ( \frac{\partial^2 {\boldsymbol \xi}}{\partial t^2} +\nabla  \Delta U -(\nabla \cdot {\boldsymbol \xi} )\nabla U_0\right ) = \nabla \cdot \Delta {\boldsymbol t}\,,
\end{align}
where $\rho_0$ is the unperturbed mass density, $U_0$ is the unperturbed gravitational potential and $\boldsymbol t$ is the stress tensor so that the right hand side represents the hydrodynamical force acting on the fluid element. In other words, the left hand side of the equation is the kinetic term and the right hand side of the equation represents the external force. Similarly, for a mass element on a two-dimensional disk, we can write down the equation of motion as
\begin{align}\label{eq:eom}
	\sigma_0 \left ( \frac{\partial^2 {\boldsymbol \xi}}{\partial t^2} +\nabla_{\parallel}  \Delta U -(\nabla_{\parallel} \cdot {\boldsymbol \xi} )\nabla U_0\right ) = {\boldsymbol F}_{\rm disk-in} +{\boldsymbol F}_{\rm disk-out} \, ,
\end{align}
where $\sigma_0$ is the unperturbed surface mass density and $\Delta U$ is the Lagrangian potential perturbation. Since we only consider the vertical displacement,  $\boldsymbol{\xi}$ is divergence-free $\nabla \cdot \boldsymbol{\xi} =0$, i.e., there are no density perturbations.
In the equilibrium case, $U_0$ satisfies $\nabla^2 U_0 =0$ except at the disk plane, where the vertical derivative is discontinuous:
\begin{align}\label{eq:bc}
	\left . \frac{\partial U_0}{\partial z} \right |_+ - \left . \frac{\partial U_0}{\partial z} \right |_- = 4 \pi \sigma_0\,.
\end{align}
The right hand side of Eq.~\eqref{eq:eom}  can be obtained by integrating  Eq.~\eqref{eq:3d} across the membrane (for details, see the Supplementary Material ). It can also be derived directly from the membrane configuration --- as we motivate here --- since it physical represents external forces. The external force is given by two components of disk forces. The in-plane component is generated by the pressure variation and the tilt of the disk plane:
\begin{align}
	{\boldsymbol F}_{\rm disk-in} = -\nabla_{\parallel} \Delta P + (\nabla_{\parallel} P \cdot \Delta {\bf n})\hat{e}_z\,,
\end{align}
where ${\bf n} = \hat{e}_z -\partial_x \xi_z \hat{e}_x -\partial_y \xi_z \hat{e}_y =\hat{e}_z +\Delta {\bf n}$ is the normal vector to the disk. The pressure perturbation is related to the density perturbation through the disk equation of state: $\Delta P/P =\Gamma_1 \, \Delta \sigma/\sigma$ \cite{poisson2014gravity}, where $\Gamma_1$ depends on the equation of state and the nature of the perturbation (e.g., adiabatic or isothermal). Therefore, the Lagrangian pressure perturbation is zero for vanishing $\Delta \sigma$. The off-plane disk force is due to the warping of the disk. If we imagine the local disk surface has a radius of curvature $R$, then the magnitude of out-of-plane force is just $2 P/R$.
For general mean curvature $\kappa$, we have
\begin{align}
	\boldsymbol{F}_{\rm disk-out} =-P \kappa \, {\bf n}, \quad \text{with} \quad \kappa =\frac{\partial^2 \xi_z }{\partial x^2}+ \frac{\partial^2 \xi_z }{\partial y^2}\,.
\end{align}

In order to compute the potential perturbation $\Delta U$, in particular, its value and derivatives on the disk plane, we make a coordinate transformation so that $z' =z -\xi_z$, with $x,y$ coordinates unchanged. The disk is mapped to the $z'=0$-plane in this new coordinate system, which is more convenient for solving the boundary value problem. Using
\begin{align}
	\frac{\partial}{\partial x'} =\frac{\partial}{\partial x} +\partial_x \xi_z \frac{\partial }{\partial z}, \quad \frac{\partial}{\partial y'} =\frac{\partial}{\partial y} +\partial_y \xi_z \frac{\partial }{\partial z}, \quad \frac{\partial}{\partial z'} =\frac{\partial}{\partial z}\,,
\end{align}
the original Laplace equation $\nabla^2 U=0$  (for $z\neq \xi_z$) becomes
\begin{align}\label{eq:laplaceu}
	\nabla'^2 U =  2 \partial_{x'}\xi_z \frac{\partial^2 U}{\partial x' \partial z'} +  2 \partial_{y'}\xi_z \frac{\partial^2 U}{\partial y' \partial z'}+\left( \partial^2_{x'}\xi_z +  \partial^2_{y'}\xi_z \right)\frac{\partial U}{\partial z'}
\end{align}
with the matching conditions that $\partial_{z'} U|_+ -\partial_{z'} U|_- =4 \pi \sigma$ and $U|_+=U|_-$. Because $\boldsymbol{\xi}$ is an infinitesimal displacement, we can write $U$ as $U_0 + U_1$, with $U_0$ satisfying $\nabla'^2 U_0=0$ together with $\partial_{z'} U_0 |_+ -\partial_{z'} U_0|_- =4 \pi \sigma$ and $U_0|_+=U_0|_-$. The solution of $U_0$ is obviously known, and $U_1$ may be obtained by solving
\begin{align}\label{eq:u1}
	\nabla'^2 U_1 =  2 \partial_{x'}\xi_z \frac{\partial^2 U_0}{\partial x' \partial z'} +  2 \partial_{y'}\xi_z \frac{\partial^2 U_0}{\partial y' \partial z'}+\left( \partial^2_{x'}\xi_z +  \partial^2_{y'}\xi_z \right)\frac{\partial U_0}{\partial z'}\,,
\end{align}
with $\partial_{z'} U_1 |_+ -\partial_{z'} U_1|_- =0$ and $U_1|_+=U_1|_-$, so that $U_1$ is completely regular in the entire spacetime. In particular, $U_1$ evaluated on the disk surface can be mapped back to $\Delta U$
with $\Delta U := U(z')-U_0(z) = U_0(z')+U_1(z')-U_0(z)$, and
$\nabla\Delta U$ is is the gravitational backreaction described in Eq.~\eqref{eq:eom}.

At this point, we consider a planar  mode with $\xi \propto e^{i {\bf k} \cdot {\bf x}}$ in the eikonal limit, that is, $|k| \gg 1$. The right hand side of Eq.~\eqref{eq:eom} is dominated by ${\bf F}_{\rm disk-out}$, which is proportional to $k^2$. On the other hand, as $U_1$ is also proportional to $e^{i {\bf k} \cdot {\bf x}}$ and the source term for $U_1$ in Eq.~\eqref{eq:u1} is dominated by the term proportional to $k^2$, we have $U_1 \propto k^0$, $\nabla' U_1 \propto k$ and $\partial_{z'} U_1 \propto k$ ($\partial_{z'} \propto k$ as $1/k$ is the only length scale in the problem). So the gravitational restoring force is subdominant compared to the anti-restoring force by the warping disk. The dispersion relation is approximately (with $\partial_t \rightarrow - i \omega$)
\begin{align}\label{eq:ants1}
	\omega^2 \approx - P/\sigma_0\, k^2,
\end{align}
which leads to exponential mode growth if $P/\sigma_0 >0$.

\vspace{0.2cm}
\noindent{\bf Relativistic case.} In the relativistic setting, we consider a model problem for compact objects with a membrane: an infinite membrane with surface mass density $\sigma$ and surface pressure $P$, which is a good approximation for perturbations of (spherical) compact objects in the eikonal limit. We discuss all the steps of the derivation of the equation of motion of the membrane perturbations. Detailed manipulations are relegated to the Supplementary Material.

If we consider the spacetime of a gravastar or a thin-shell wormhole, the metric can be expressed as ${\rm diag}[-f(r), 1/h(r), r^2, r^2\sin^2\theta]$, with different prescriptions for $f(r)$ and $h(r)$. As we focus on perturbations of small wavelength, we can zoom in on the neighborhood of any point on the membrane, and rewrite the metric as
\begin{align}\label{eq:bac}
	ds^2& =g^{(0)}_{\,\mu\nu} dx^\mu dx^\nu \nonumber \\
	&=-U(z)dt^2+U_z(z)dz^2+ U_{\rm p}(z)(dx^2+dy^2) \, ,
\end{align}
where $x =\theta \, \cos\phi$ and $y = \theta \, \sin\phi$. This local representation of the membrane metric is generic. The Israel boundary conditions on the membrane relate the extrinsic curvature $K_{ij}$ to the surface-layer property by  \cite{Poisson:1995sv} ($d \tau = \sqrt{U}dt$)
\begin{align}
	K^x{}_x |^+_- & = K^y{}_y |^+_- = \left. \frac{1}{\sqrt{U_z}}\frac{U_{\rm p}'}{2U_{\rm p}} \right |^+_-=-4 \pi \sigma \,,\nonumber \\
	K^\tau{}_\tau |^+_- & =\left. \frac{1}{\sqrt{U_z}}\frac{U'}{2U}  \right |^+_-= 8 \pi (P+\sigma/2)
\end{align}
where $|^+_-$ indicates the difference between $0_+$ and $0_-$ of the membrane in the $z$-direction. Since we can always rescale $z$ in the vertical/radial direction, in the rest of the discussion we shall set $U_z=1$.

Let us now assume the membrane is perturbed with vertical displacement $\xi_z=\xi(x,y,t)$. The membrane stress energy tensor is given by
\begin{align}
	\tau^{\mu\nu} = \delta(z-\xi) \left [ (\sigma+P) u^\mu u^\nu+P (g^{\mu\nu} -n^\mu n^\nu)\right ]\,,
\end{align}
where ${\bf n} $ is the normal vector of the membrane.  It is given by $e_z(1-h_{zz}/2)
-\sum_{\alpha=t,x,y} (\xi_{,\alpha}+h_{\alpha z})e_\alpha/g^{(0)}_{\alpha\alpha}$ and $e_x$ is given by $\tfrac{\partial}{\partial x}$ (similarly for $e_y$ and $e_z$), where $h_{\mu\nu}$ is sourced by the membrane motion (compare with the right hand side of Eq.~\eqref{eq:laplaceu}). In order to derive the equation of motion for $\xi_z$, we  transform to the coordinate system with $z'=z-\xi, t'=t,x'=x,y'=y$, such that the membrane is mapped back to the ``equatorial'' plane in the new coordinates.
The spacetime metric in the new coordinates can be written as $g_{\mu\nu} = g^{(0)}_{ \mu\nu} +\xi_{\mu | \nu}+\xi_{\nu | \mu}+h_{\mu\nu} = g^{(0)}_{ \mu\nu}+\tilde{h}_{\mu\nu}$, where $|$ represents the covariant derivative with respect to $g^{(0)}_{\mu\nu}$. The gravitational perturbation is more conveniently computed in the original $(t,z,x,y)$ coordinate system:
\begin{align}\label{eq:wave}
	\bar{h}_{\mu\nu | \alpha}{}^\alpha +2 R_{\alpha \mu\beta \nu} \bar{h}^{\alpha \beta} =0\,,
\end{align}
with the trace-reversed $\bar{h}_{\mu\nu} := h_{\mu\nu}-\tfrac{1}{2} h \,g^{(0)}_{\mu\nu}$ and assuming the Lorenz gauge condition $\bar{h}_{\mu\alpha}^{\quad | \alpha}=0$, which is preserved along the evolution driven by the wave equations if it is initially satisfied. The waves should be outgoing at infinity and the matching condition at the membrane leads to
\begin{align}\label{eq:pm}
	\int^+_- dz \; \bar{h}_{\mu\nu | \alpha}{}^\alpha = \bar{h}_{\mu\nu,z}|^+_-=8 \pi \, \delta \tau_{\mu\nu}
\end{align}
with $\delta \tau^{xz} = P \partial_x \xi/U_{\rm p}, \delta \tau^{yz} = P \partial_y \xi/U_{\rm p},
\delta \tau^{tz} =\sigma\dot{\xi}/U $. The metric functions are continuous across the membrane, and only $\partial_z h_{\mu\nu}$ may be discontinuous (we shall assume a simple setup with reflection symmetry, where $\partial_z h_{\mu\nu} |_-+\partial_z h_{\mu\nu} |_+=0$, but the final result does not rely on this assumption).
In the eikonal limit, $\partial_t, \partial_x, \partial_y, \partial_z$ all scale as $k$, which suggests that the boundary value for $\bar{h}_{\mu\nu} =\mathcal{O}(k)^0$ and $\partial \bar{h}_{\mu\nu} = \mathcal{O}(k)$. Their interior value should have similar scaling laws following the wave equation in Eq.~\eqref{eq:wave}. (Such coupled wave equations in Lorenz gauge can be solved numerically in Schwarzschild spacetime \cite{Barack:2005nr}, or perturbatively with WKB method because the separation of scales in $1/k$ and the curvature radius of the background spacetime $\sim U/U'$.)

The  equation of motion for $\xi$ is given by $T^{z'\nu} {}_{; \nu}=0$. We integrate it from lower side to the upper side of the membrane ($z'=0_- \rightarrow 0_+$), which becomes (evaluated at $z'=0$)
\begin{align}
	(\sigma+P)(u^{z'} u^\nu)_{;\nu} = P (n^{z'} n^\nu)_{;\nu}
\end{align}
or, more explicitly,
\begin{align}
	\frac{\sigma+P}{2 U} (2\tilde{h}_{tz',t}-\tilde{h}_{tt,z'}) =&-\frac{P}{2 U} \tilde{h}_{tt,z'} +\frac{P }{2U_{\rm p} }(\tilde{h}_{xx,z'}+\tilde{h}_{yy,z'}) \nonumber \\
	&+\frac{P\tilde{h}_{tz',t}}{U}-\frac{P\tilde{h}_{xz',x}}{U_{\rm p}}-\frac{P\tilde{h}_{yz',y}}{U_{\rm p}} \, .
\end{align}
By noticing that $\tilde{h}_{\mu\nu} =h_{\mu\nu}+\xi_{\mu|\nu}+\xi_{\nu | \mu}$ and $h_{\mu\nu, z'}|_++h_{\mu\nu,z'} |_-=0$, the equation reduces to
\begin{align}
	\frac{\sigma}{ U} \xi_{,tt}+\frac{P}{U_{\rm p}}(\xi_{,xx}+\xi_{,yy})
	=-\frac{\sigma}{ U} h_{tz,t}-\frac{P h_{xz,x}}{U_{\rm p}}-\frac{P h_{yz,y}}{U_{\rm p}} \, .
\end{align}
It is clear that the $\xi_{,xx}+\xi_{,yy}$ terms here provide the anti-spring force that potentially drives the instability. However, to fully address the mode dispersion relation, we also need to account for the gravitational backreaction. The relevant terms in the eikonal limit are described by the terms on the right-hand side, which all scale as $k$ according to the discussion under Eq.~\eqref{eq:pm}. Therefore similar to the Newtonian case,
the relativistic anti-spring force effect scales as $k^2$ and gravitational backreaction scales as $k$. In the eikonal limit, we therefore find
\begin{align}\label{eq:instab}
	\omega^2 \approx - \frac{U(0) P}{U_{\rm p}(0)\sigma}\, k^2\,,
\end{align}
which signals an instability if $P/\sigma>0$. This result can be straightforwardly extended to cases for which the upper and lower spacetime have different cosmological constants.

\vspace{0.2 cm}

\noindent{\bf Gravastars.} A gravastar can be modeled by a spherical membrane separating an inner de Sitter spacetime and an outer Schwarzschild spacetime. If  the inner region is an anti-de Sitter (AdS) spacetime, it is usually called an AdS Bubble \cite{Danielsson:2017riq,Danielsson:2021ykm}. Defining $\rho$ as the ``energy density" or cosmological constant in the inner space, $a$ as the radius of the membrane, $\sigma$ as the membrane surface energy density and $P$ as its pressure, the total mass $M$ of the spacetime is (following the notation in \cite{Pani:2009ss})
\begin{align}\label{eq:mmv}
	M = M_v + M_s \sqrt{1-\frac{2 M_v}{a}} +\frac{M^2_s}{2 a}\, ,
\end{align}
where $M_s = 4 \pi a^2 \sigma$ is the thin-shell mass and $M_v =4 \pi \rho a^3/3$ is the volume energy within the shell. The pressure within the shell is related to these masses through
\begin{align}\label{eq:p}
	P = & \frac{1}{8 \pi a} \left [- \frac{1-4 M_v/a}{\sqrt{1-2 M_v/a}} +\frac{1-M/a}{\sqrt{1-2M/a}}\right ]\nonumber \\
	=& \frac{1}{8 \pi a} \left [\frac{3 M_v/a}{\sqrt{1-2 M_v/a}}- \frac{1- M_v/a}{\sqrt{1-2 M_v/a}} +\frac{1-M/a}{\sqrt{1-2M/a}}\right ]\,.
\end{align}
To ensure meaningful values for $P$, we require that $M/a \le 1/2$ and $M_v/a \le 1/2$. If the gravastar satisfies the weak energy condition, the surface density $\sigma$ and $M_s$ are both positive. We notice that $M\ge M_v >0$ according to Eq.~\eqref{eq:mmv} (note $(1-x)/\sqrt{1-2x}$ is a monotonically increasing  for $0 \le x \le 1/2$).
From the second line of Eq.~\eqref{eq:p}, it is straightforward to see that the pressure is always positive. Intuitively it can be viewed as a consequence of the outer spacetime squeezing the inner spacetime, as the outer spacetime has larger effective pressure than the inner spacetime (also with the self-gravitation of the membrane). Although the analysis in the previous section was with topology $\mathbb{R}^2$ while the membrane of gravastars has topology $\mathbb{S}^2$, this distinction is irrelevant as we consider local perturbations in the eikonal limit. This simple observation, together with the analysis of the warping mode instabilities, immediately suggests that {\it  gravastars satisfying the weak energy condition are unstable}.
The instability timescale is determined by Eq.~\eqref{eq:instab} and depends on the prescription for $P$ and $\sigma$.


\begin{figure}
				\includegraphics[width=0.48\textwidth]{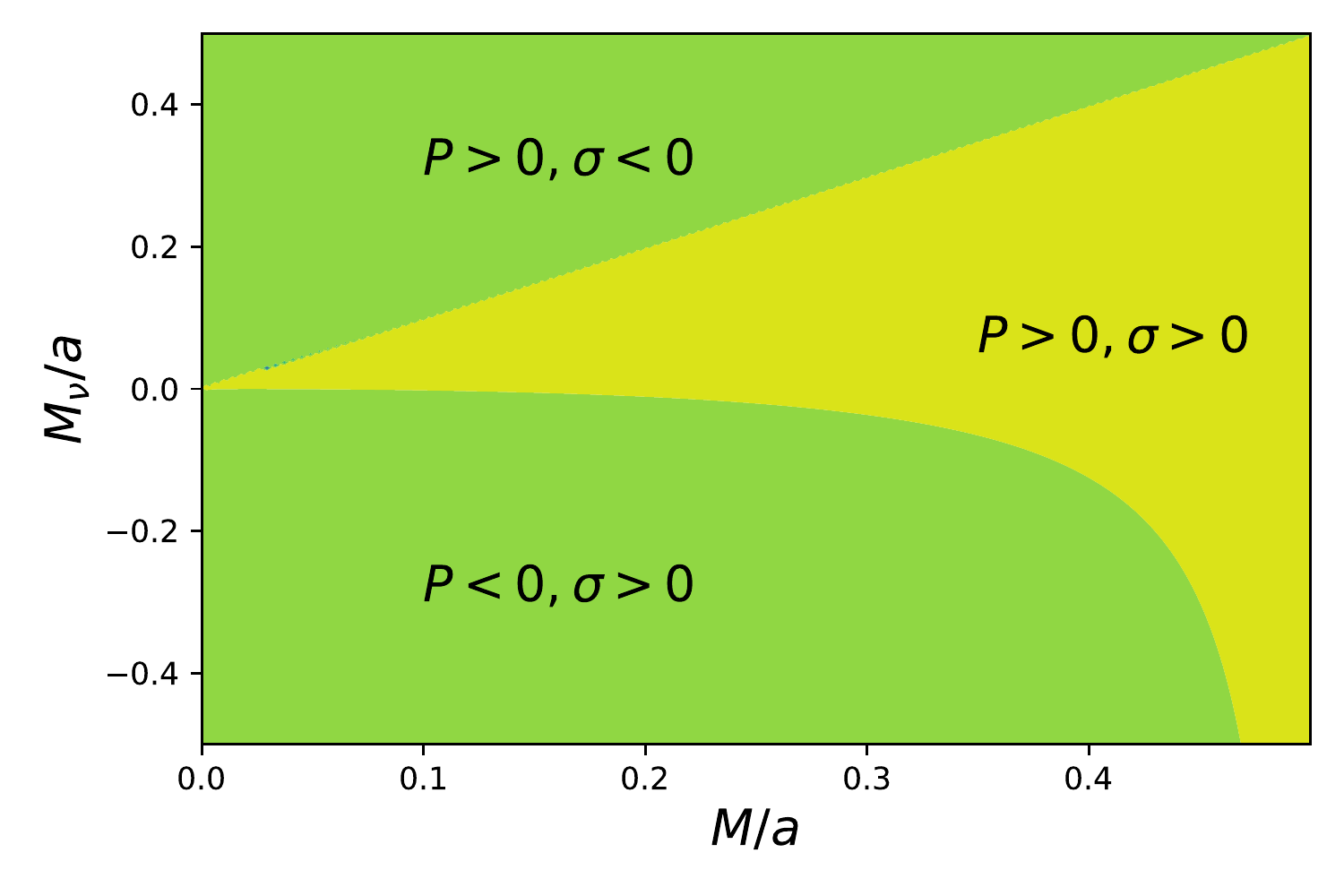}

			\caption{The ``phase diagram" of gravastars (with $M_v >0$) and AdS Bubbles (with $M_v<0$). The regime with positive pressure and density is associated with the warping instability.}
			\label{fig:pplot}
		\end{figure}

In the more general setting, as we consider both de Sitter and AdS interiors and surface density with arbitrary sign, the warping instability applies part of the parameter space of gravastars and AdS Bubbles, as shown in Fig.~\ref{fig:pplot}.

The modal stability of gravastars was initially studied in \cite{Pani:2009ss}, which explicitly computed the quasinormal mode frequency for $\ell =2$ axial and polar perturbations. However, the analysis in \cite{Pani:2009ss} treats the membrane as the provider of the matching condition between the inner and outer spacetime, in the same spirit as Eq.~\eqref{eq:bc}, but did not incorporate the membrane oscillations into the coupled mode equations. An explicit discussion of the gravastar mode analysis is included in the Supplementary material.  It is indeed the membrane modes that destabilize the whole system in the eikonal limit.

\vspace{0.2cm}
\noindent{\bf Thin-shell wormholes.}
There are other horizonless compact objects generally considered in the literature as BH mimickers, or as candidates sourcing gravitational wave echoes.
For example, thin-shell wormholes are commonly studied objects with compactness arbitrarily close to a BH. Consider two Schwarzschild solutions of the same mass $M$ attached at radius $r_0$  \cite{Poisson:1995sv,Cardoso:2016rao}, the corresponding thin-shell pressure and density at the wormhole throat are
\begin{align}
	P =\frac{1}{4 \pi r_0} \frac{1-M/r_0}{\sqrt{1-2M/r_0}} \, , \quad \sigma =-\frac{1}{2\pi r_0}\sqrt{1-2M/r_0}
\end{align}
so that the pressure  is positive and the density is negative, which implies that static thin-shell wormholes are free from the warping instability. On the other hand, the empty shell models (which have positive surface energy density) as considered in \cite{Cardoso:2016oxy,Bonga:2021ouq} naturally require positive in-shell pressure to support against gravity, and are all unstable to warping perturbations in the eikonal limit \footnote{The purpose of \cite{Bonga:2021ouq} is to theoretically demonstrate that a compact object can have the same set of multipole moments of Kerr, instead of proposing its astrophysical relevance.}.

\vspace{0.2cm}
\noindent{\bf Discussion.}
We have discovered a generic instability for thin-layer structures in General Relativity, if $P/\sigma>0$, with applications highlighted in exotic compact objects with membranes.
One may imagine various ways to ``cure" these systems so that they are free from warping instabilities. One possible way is to
add additional rigidity against warping for the membrane, e.g, a new term in the action with
\begin{align}
	S = \alpha \int d^3 \xi \; K^{ij} K_{ij}
\end{align}
where $\xi$ is the parametrization for the ``world tube" of the membrane,  $K^{ij}$ is the extrinsic curvature and $\alpha$ is a positive constant characterizing the rigidity. A possible caveat is that such an additional term in the action may lead to higher-order derivative terms in the equation of motion, which may raise concerns regarding well-posedness of the problem.
Moreover, adding dissipation to the system does not cure the instability. This is because the anti-spring force causes run-away behavior of the displacement instead of oscillations. If the displacement were to saturate at some value, the dissipation becomes zero as there is zero velocity, but the anti-spring force continues to drive the displacement to larger values, i.e., there is no saturation point. On the other hand, if we replace the membrane with a shell of matter of thickness $d$, this can remove the instability. The thickness essentially adds a spatial frequency cutoff $k \sim 1/d$ in the above analysis. The caveat is that $d$ has to be sufficiently large so that the anti-spring in Eq.~\eqref{eq:ants1} becomes sub-dominant. Note that in this case the description for the dynamic behaviour of matter with anisotropic stress is highly nontrivial and currently unknown.

A membrane with  density and pressure having the same sign generically prefers configurations with higher surface curvature as they are associated with a lower energy state, if gravitational backreaction is neglected. For example, a membrane with an ellipsoidal shape has lower potential energy than that with a spherical shape. Mathematically the potential energy is $\propto Q^{2}_{ij}/(2 \lambda)$ where $Q_{ij}$ is the mass quadrupole moment and $\lambda$ the tidal Love number. Negative potential energy means that $\lambda$ is negative. Even with gravitational backreaction included, if it is weaker than the anti-spring force such that the potential energy is still negative, the Love number $\lambda$ will also be negative \footnote{Private communications with Eric Poisson.}. Therefore the warping instability is connected to the negativity of tidal Love numbers, which applies to generic deformations with any $\ell \ge 2$ \cite{Pani:2015tga}. In the eikonal limit, the tidal Love number $\lambda_\ell$ has to be negative if $P/\sigma>0$.

\vspace{0.2cm}
{\it Acknowledgements---}
We thank Luis Lehner, Eric Poisson and Yanbei Chen for helpful discussions.
HY and ZP are supported by the Natural Sciences and Engineering Research
Council of Canada and in part by Perimeter Institute for Theoretical Physics.
Research at Perimeter Institute is supported in
part by the Government of Canada through the Department of
Innovation, Science and Economic Development Canada and
by the Province of Ontario through the Ministry of Colleges
and Universities.

\newpage
\section{Self-gravitating membrane in the Newtonian setting}

The treatment of membrane motion in the Newtonian limit offers an intuitive example for various driving terms in the membrane's equation of motion, especially the dominant disk force and the gravitational back-reaction. By integrating the fluid variables from $0_-$ side to the $0_+$ side of the membrane, it is straightforward to obtain the left hand side of Eq.~(3) in the main text.
Here, we will present a derivation for the right hand side of Eq.~(3), followed by explanations for the physical meaning of the different parts.

\begin{figure}
			\includegraphics[width=0.5\textwidth]{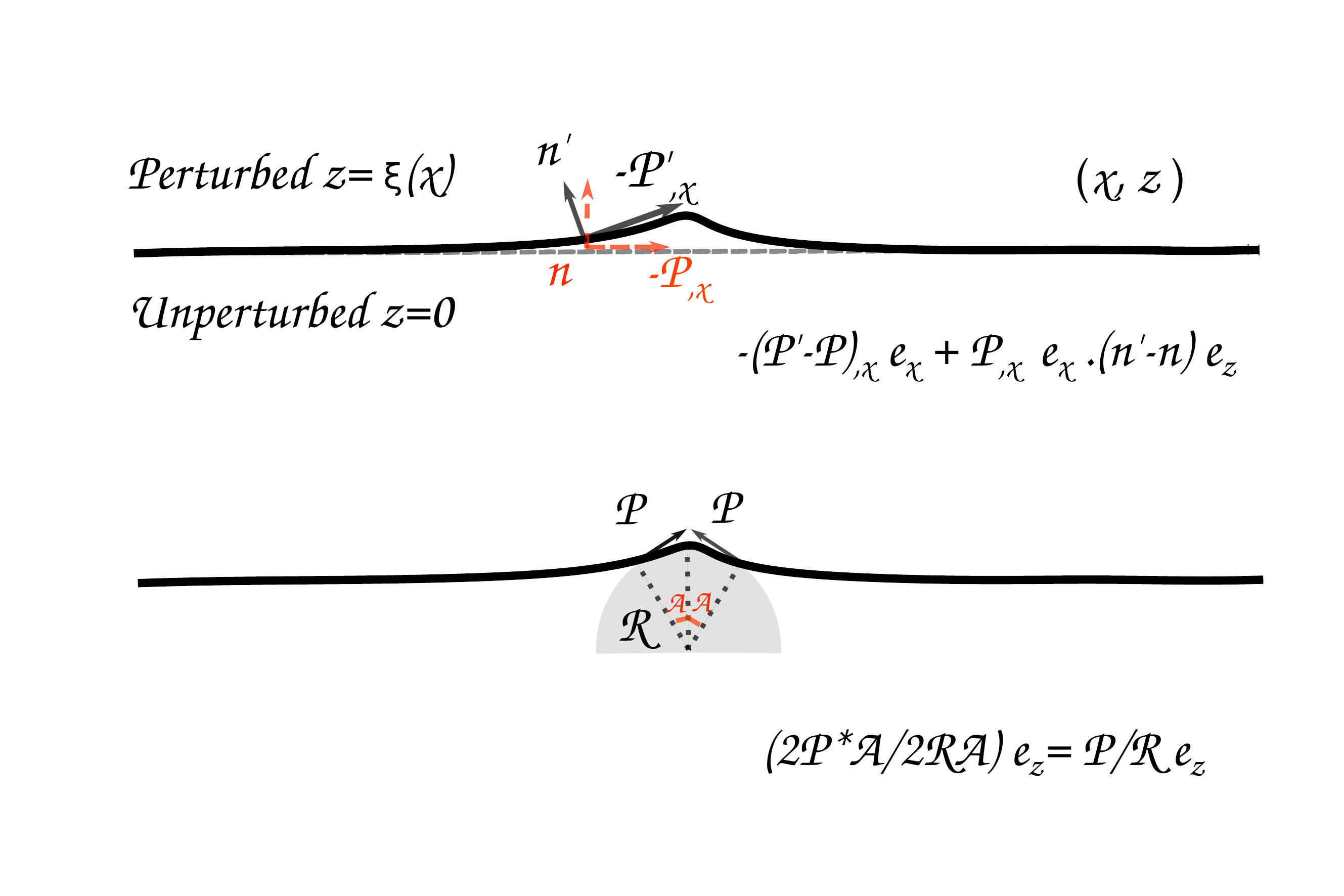}
			\caption{This figure illustrates the physical meaning of ${\bf F}_{\rm disk-in}$ (top) and ${\bf F}_{\rm disk-out}$ (bottom).}
			\label{fig:app1}
		\end{figure}

The anisotropic stress of the membrane can be written as
\begin{align}
{\bf t}_{ij} = -P (\delta_{ij}-n_i n_i) \delta[z-\xi(t)]
\end{align}
where ${\bf n}$ is the normal vector of the membrane.
After performing the average from the bottom to the top side of the membrane, the $\delta(z-\xi)$ function in $\Delta {\bf t}$ is removed and $\nabla \cdot \Delta {\bf t}$ becomes $\nabla_{\parallel} \cdot \langle \Delta {\bf t} \rangle ^+_-$, where $\nabla_{\parallel} =\nabla-{\bf n}({\bf n} \cdot \nabla)$ is the gradient within the surface of the membrane.
With nonzero $\boldsymbol{\xi}$, the normal vector becomes
\begin{align}
{\bf n} \longrightarrow \hat{e}_z+\Delta {\bf n} =\hat{e}_z -\partial_x \xi \hat{e}_x -\partial_y \xi \hat{e}_y
\end{align}
so that
\begin{align}
\nabla_{\parallel} =\partial_x \hat{e}_x+\partial_y \hat{e}_y + (\partial_x \xi \partial_x +\partial_y \xi \partial_y) \hat{e}_z
	+ (\partial_x \xi \hat{e}_x +\partial_y \xi \hat{e}_y) \partial_z\,.
\end{align}
The term $(\partial_x \xi \partial_x+\partial_y \xi ) \hat{e}_z \cdot {\bf t}$ in $\nabla_{\parallel} \cdot  \Delta {\bf t}$ is $\mathcal{O}(\xi^2)$ and therefore does not contribute at linear order. The $(\partial_x \xi \hat{e}_x +\partial_y \xi \hat{e}_y) \partial_z \cdot  \Delta {\bf t}$ term also does not contribute because after performing the average across the vertical direction of the membrane $\langle \Delta {\bf t} \rangle ^+_-$ has no explicit dependence on $z$, {i.e., $\nabla_{\parallel}\cdot \Delta {\bf t} =(\partial_x \hat{e}_x+\partial_y \hat{e}_y)\cdot \Delta {\bf t}$.} Therefore, we find that
\begin{widetext}
\begin{align}
\nabla_{\parallel} \cdot \langle \Delta {\bf t} \rangle ^+_- \longrightarrow -\nabla_{\parallel,i} \left \{ (P_0+\Delta P) \left [\delta_{ij}-(\hat{e}_{z,i}+\Delta n_i) (\hat{e}_{z,j}+\Delta n_j) \right ] 
-P_0 \left [\delta_{ij}-\hat{e}_{z,i}\hat{e}_{z,j} \right ]\right \} \; ,
\end{align}
\end{widetext}
where $\nabla_{\parallel,i}$ can be thought of as the unperturbed derivative operator along the membrane.
This is equal to
\begin{align}
-\nabla_{\parallel } \Delta P + (\nabla_{\parallel} P \cdot {\Delta n} )\hat{e}_z - P_0 (\partial^2_x \xi+\partial^2_y \xi) \hat{e}_z ={\bf F}_{\rm disk-in} +{\bf F}_{\rm disk-out}\,.
\end{align}
Let us now proceed to discuss the physical meaning of different terms. For simplicity, we can neglect the $y$-dependence and visualize the two forces in Fig.\ref{fig:app1}. ${\bf F}_{\rm disk-in}$ is shown in the top panel, with its first term corresponding to the additional force generated by the pressure perturbation and the second term is produced by the tilt of the orbital plane, so that the pressure force $\nabla P$ points to a slightly different direction. ${\bf F}_{\rm disk-out}$   is shown in the bottom panel and is generated by the warping of the membrane. For example, in this effective one-dimensional scenario for Fig.\ref{fig:app1}, $\partial^2_x \xi = -\xi/R$, with $R$ being its local curvature radius. It is straightforward to see that the net force is along the $z$-direction, with force density $P/R$ as consistent with the expression for ${\bf F}_{\rm disk-out}$.

\section{Self-gravitating membrane in the relativistic setting}

In this section, we will show how the equation of motion for $\xi$ in Eq.~(18) in the main text derives from Eq.~(16), which we repeat here for convenience:
\begin{align}
(\sigma+P)(u^{z'} u^\nu)_{;\nu} = P (n^{z'} n^\nu)_{;\nu}\ . \label{eq:conservation}
\end{align}

Let us start by expanding the left hand side, but first note that in $(t,z',x,y)$ coordinates $g_{\mu\nu} = g^{(0)}_{\mu\nu} + \xi_{\mu|\nu}+\xi_{\nu|\mu}  +  h_{\mu\nu} = g^{(0)}_{\mu\nu} +\tilde h_{\mu\nu}$ where $g^{(0)}_{\mu\nu} = {\rm diag}\{-U(z), 1, U_p(z), U_p(z)\}$, $\xi_\mu=(0, \xi(t,x,y), 0, 0)$
and
$u^\mu = (u^t, u^{z'}, u^x, u^y) = (u^t, 0, 0, 0)$, where $u^t$ is constrained by its normalization. In particular, $u^t$ accurate to the linear order is
\begin{align}
	u^t
	&=\frac{1}{\sqrt{U(z)-\tilde h_{tt}}} = \frac{1}{U^{1/2}} + \frac{1}{2}\frac{\tilde h_{tt}}{U^{3/2}} \; .
\end{align}
Substituting these expressions, we obtain
\begin{align}
	u^{z'}_{;\nu} u^\nu = u^{z'}_{;t} u^t = (u^{z'}_{,t} + u^\gamma \Gamma ^{z'}_{\gamma t})u^t = u^tu^t  \Gamma ^{z'}_{tt}
	\overset{\mathcal{O}(\tilde h)}{=} \frac{1}{U}\left(\tilde h_{tz',t} -\frac{1}{2}\tilde h_{tt,z'} \right)
\end{align}
and $u^{z'}u^\nu_{;\nu}\overset{\mathcal{O}(\tilde h)}{=} 0$,  where we have used the reflection condition
$U_{,z'}|_{+}+U_{,z'}|_{-}=0$.
Therefore, the left hand side of Eq.~\eqref{eq:conservation}, up to linear order in the perturbations, is given by
\begin{align}
(\sigma+P)(u^{z'} u^\nu)_{;\nu}  \overset{\mathcal{O}(\tilde h)}{=}    \frac{\sigma+P}{U}\left(\tilde h_{tz',t} -\frac{1}{2}\tilde h_{tt,z'}\right) \; .
\end{align}

For expanding the right hand side, we start with the normal vector $n$ that is perpendicular to the surface $z'=0$ so that $n_\mu \propto (0, 1, 0, 0)$ and
\begin{align}
	n^\mu = (n^t, n^{z'}, n^x, n^y) =
	\left(\frac{\tilde h_{tz'}}{U}, 1-\frac{1}{2}\tilde h_{z'z'}, -\frac{\tilde h_{xz'}}{U_p},
	-\frac{\tilde h_{yz'}}{U_p} \right) \ ,
\end{align}
accurate to the linear order,
where we have normalized the vector $g_{\mu\nu}n^\mu n^\nu = 1$.
Then it is straightforward to show that
\begin{align}
	n^\nu_{;\nu} &= n^\nu_{,\nu} + n^\gamma \Gamma^\nu_{\gamma\nu}
	= (n^{z'}_{,z'} + n^{z'}\Gamma^\nu_{z'\nu}) + (n^t_{,t}+n^x_{,x}+n^y_{,y} + n^t\Gamma^\nu_{t\nu}
	+ n^x\Gamma^\nu_{x\nu} + n^y\Gamma^\nu_{y\nu})\nonumber\\
	&\overset{\mathcal{O}(\tilde h)}{=}
 \frac{1}{2}\left(-\frac{\tilde h_{tt,z'}}{U} + \frac{\tilde h_{xx,z'}+\tilde h_{yy,z'}}{U_p} \right)
	+\left(\frac{\tilde h_{tz',t}}{U}-\frac{\tilde h_{xz',x}+\tilde h_{yz',y}}{U_p} \right)
\end{align}
and $n^\nu n^{z'}_{;\nu} 	\overset{\mathcal{O}(\tilde h)}{=} 0$. Consequently, the right hand side of Eq.~\eqref{eq:conservation} is
\begin{align}
 P (n^{z'} n^\nu)_{;\nu} =  -\frac{P}{2U}\tilde h_{tt,z'} + \frac{P}{2U_p} (\tilde h_{xx,z'} + \tilde h_{yy,z'})
+\frac{P}{U}\tilde h_{tz',t} -\frac{P}{U_p}(\tilde h_{xz',x} + \tilde h_{yz',y})
\end{align}

Equating both sides, we obtain
\begin{align}
	\frac{\sigma}{U}( h_{tz,t}+ \xi_{,tt}) +\frac{\sigma}{U} h_{tt,z'} = &\frac{P}{2U_p}(h_{xx,z'}+h_{yy,z'}) \nonumber \\
	&-\frac{P}{U_p} (h_{xz,x} +\xi_{,xx} + h_{yz,y} + \xi_{,yy})\ ,
\end{align}
where we have used $\tilde h_{\mu\nu} = h_{\mu\nu}+\xi_{\mu|\nu}+\xi_{\nu|\mu}$.
Finally, using the reflection condition $h_{\mu\nu,z'}|_+ + h_{\mu\nu,z'}|_-=0$,
the above equation reduces to Eq.~(18) in the main text:
\begin{align}
	\frac{\sigma}{U} \xi_{,tt} + \frac{P}{U_p}(\xi_{,xx} + \xi_{,yy})
	= -\frac{\sigma}{U} h_{tz,t} - \frac{P}{U_p}(h_{xz,x}+h_{yz,y})\,.
\end{align}

\section{Explicit application to gravastar perturbations}

The derivation for the warping instability of thin-layers applies for general settings. To emphasize and illustrate its application for black hole mimickers, we discuss non-radial oscillations of gravastars, as initially presented in  Ref.~[37] of the main text, but now with the membrane motion properly taken into account. This example should provide intuition for the warping instability.

Let us consider a gravastar with the thin-shell located at $r=a$ with surface density $\sigma$ and surface pressure $P$. After performing the spherical decomposition, we focus on even partity/polar metric perturbations in the interior and exterior regime of the gravastar for which the non-zero metric perturbations are
\begin{align}
h_{tt} &=f(r) H(t,r) Y_{\ell m}, h_{t r} =H_1(t,r) Y_{\ell m}, h_{rr} = H_2(t,r)/h(r) Y_{\ell m}\,,\nonumber \\
h_{\theta \theta} & = r^2 K(t,r) Y_{\ell m}, h_{\phi\phi} = r^2\sin^2\theta K(t,r) Y_{\ell m}\,.
\end{align} 
The functions $f(r), h(r)$ are $f(r)=\alpha h(r) =\alpha (1-2M_v r^2/a^3)$ in the interior and $f(r)=h(r)=1-2M/r$ in the exterior, with $\alpha = (1-2M/a)/(1-2M_v/a)$. At any time slice, the even-parity displacement of the membrane for a given spherical harmonic index $(\ell,m)$ can be written as
\begin{align}
\xi_{(1)}^\alpha = [0, h(r)z(t) Y_{\ell m},0,0],\quad \xi_{(2)}^\alpha =[0,0,\frac{w(t)}{r^2}Y_{\ell m,\theta}, \frac{w(t)}{r^2\sin^2\theta} Y_{\ell m, \phi}]\,,
\end{align}
for two independent vector spherical harmonic basis. To derive the matching conditions and the equations of motion of the membrane, it is convenient to perform a coordinate transformation to map each mass element to its original coordinate value, as implemented in Ref.~[37]. So in the new coordinate we have $\tilde{h}_{\mu\nu} =h_{\mu\nu} +h^{(\xi)}_{\;\mu\nu}, \tilde{g}_{\mu\nu} = g_{0 \mu\nu}+\tilde{h}_{\mu\nu}$, where $h^{(\xi)}_{\;\mu\nu}$ is
\begin{align}
	h^{(\xi)}_{\;\mu\nu}=
&\begin{bmatrix}
-f' h z & \dot{z} & 0 &  0\\
* &  h' z/h & z \partial_\theta &  z \partial_\phi \\
0 & * & 2 r h z & 0 \\
0 & * & 0 & 2 r h z \sin^2\theta
\end{bmatrix} Y_{\ell m} \nonumber \\
&+\begin{bmatrix}
0 & 0& \dot{w} Y_{\ell m,\theta} &  \dot{w} Y_{\ell m,\phi}\\
0 &  0 & -\frac{2 w}{r} Y_{\ell m,\theta} & -\frac{2 w}{r} Y_{\ell m,\phi}\\
* & * & 2 w Y_{\ell m|\theta \theta} & 2 w Y_{\ell m|\theta \phi} \\
* & * & * & 2 w Y_{\ell m|\phi\phi}
\end{bmatrix} \,,
\end{align}
 with $*$ being the nonzero symmetry elements and $|$ the covariant derivative on the 2-sphere. 

In the new coordinates, as discussed in Ref.~[37], the Israel junction conditions imply that
\begin{widetext}
\begin{align}\label{eq:ma}
&[\sqrt{h} z]^+_- =0, \quad K^+_- =8 \pi \sigma_0\sqrt{h} z,\quad H^+_- =8\pi (\sigma_0+2 P_0) \sqrt{h} z \,,\nonumber \\
& \left [ \sqrt{h}\left ( \frac{H}{a} -K' \right )\right ]^+_-+\left[ \frac{2 h}{a^2} -\frac{h'}{a}\right ]^+_- \sqrt{h} z =8 \pi \delta \sigma\,, \nonumber \\
& \left [ \sqrt{h} \left( K'-H'+\frac{2 \dot{H_1}}{f}\right )\right ]^+_- -\left [ \sqrt{h} \left ( 1+\frac{a f'}{2 f}\right ) \frac{H}{a}\right ]^+_- +\left [\frac{h'}{a} -\frac{2h}{a^2}+\frac{f'' h}{f}-\frac{f' h'}{2 f}\right ]^+_- \sqrt{h}z =16 \pi \delta P\,.
\end{align}
\end{widetext}
At this point, it was assumed in Ref.~[37] that both the interior and exterior metric perturbation satisfy the Regge-Wheeler gauge condition, so that the metric quantities are all functions of the master variable $\Psi, \Psi'$. As a result, one only needs two independent matching conditions, which can be directly obtained from the above equations after imposing $\delta P =v^2_s \; \delta \sigma$ (where $v_s$ can be interpreted as the speed of sound on the thin shell). With the matching conditions implemented, the metric quantities in the entire spacetime is solved. Finally, the resulting metric can then be plugged back into Eq.~\eqref{eq:ma} to obtain the values of $z, \delta \sigma, \delta P$, which are consequently no longer independent quantities.

This observation already hints at an inconsistency in the  above analysis, as the membrane can also be assumed to have certain prescribed motion (e.g., with prescribed amplitude and frequency), and the metric perturbations in the spacetime should respond to the prescribed membrane motion by solving the wave equation with the Israel matching conditions. In other words, there are infinite ways to  construct $z, \delta P, \delta \sigma$ and find the corresponding metric quantities, as we have not specified the membrane equation of motion in the above analysis. Indeed this inconsistency comes from the Regge-Wheeler gauge condition. It is allowed to impose this gauge condition in both the interior and exterior space, but then the metric perturbations will not necessarily satisfy the  continuity conditions as required in Eq. (A14)-(A20) in Ref.~[37]. Put differently, if we were to impose the Regge-Wheeler gauge in the interior and use the matching conditions to compute the value and the derivatives of the metric quantities on the exterior side of the membrane, the resulting exterior perturbations generally do not satisfy the Regge-Wheeler gauge condition.

An appropriate procedure to compute the quasinormal mode can be that imposing the Regge-Wheeler gauge in the interior space, writing down the matching conditions and the membrane equations of motions (which give the prescription for $z,w,\delta P, \delta \sigma$). After that the value and derivative  of metric quantities on the upper surface of the membrane is known, one can compute the value and derivative of the master variable $\Psi$, which is gauge invariant. Finally by solving the wave equation of $\Psi$ and imposing the outgoing condition at infinity, the quasinormal mode frequency can be determined.

For our discussion here, as we want to reveal the existence of the warp instability in the eikonal limit $\ell \gg 1$, a few observations can be made. First of all, as we focus on the radial oscillation mode, for reasons that will become clear later, the amplitude of $w$ is $1/\ell$ times smaller than the amplitude of $z$, i.e., $|w| \sim |z|/\ell$. As the perturbation of the surface density is expected to be $\delta \sigma \propto \nabla_{||} \cdot \xi \propto \ell w$, and $\delta P \propto v^2_s \delta \sigma$, we find the following scaling $\delta P, \delta \sigma \sim \mathcal{O}(\ell)^0 z$. The matching condition in Eq.~\eqref{eq:ma} implies that $H, K, H_1 \propto \mathcal{O}(\ell)^0 z $ as well.
Second, we can write down the membrane equation of motion 
\begin{align}\label{eq:eom}
\left [ (\sigma_0+\delta \sigma +P_0+\delta P) u^\mu u^\nu \right ]_{;\nu} = \left [ (P_0+\delta P) (n^\mu n^\nu -g^{\mu\nu})\right ]_{;\nu}
\end{align}
where $n_\nu\propto (0,1,0,0)$ and $u^\nu = (1/\sqrt{-\tilde{g}_{tt}},0,0,0)$. The equation of motion for $\tilde{z} := \sqrt{h}z$ (which is continuous across the membrane) is obtained from the $\mu=r$ component:
\begin{align}
&\frac{4 h}{r} \tilde{h}_{rr} +\frac{2 \cot\theta}{r^2} \tilde{h}_{r \theta} +\frac{2}{r^3} \left( \tilde{h}_{\theta \theta}+\frac{\tilde{h}_{\phi\phi}}{\sin^2\theta}\right ) - \frac{\sigma_0}{P_0} \frac{h f'}{f} \tilde{h}_{rr} + \frac{\sigma_0}{P_0}\frac{f'}{f^2} \tilde{h}_{tt} \nonumber \\
&+\frac{2}{r^2\sin^2\theta } \tilde{h}_{r \phi,\phi} +\frac{2}{r^2 } \tilde{h}_{r \theta,\theta} -\frac{1}{r^2} \left ( \tilde{h}_{\theta\theta,r}+\frac{1}{\sin^2\theta} \tilde{h}_{\phi\phi,r}\right ) + \frac{\sigma_0}{P_0} \frac{2}{f} \tilde{h}_{tr,t} \nonumber \\
& - \frac{\sigma_0}{P_0} \frac{\tilde{h}_{tt,r}}{f}  - \frac{4}{r P_0} \delta P + \frac{f'}{f \, P_0} \delta \sigma =0 
\end{align}
where an average over plus and minus side of the membrane is implicitly performed. By assigning the derivatives an order $\mathcal{O}(\ell)$, we can read of the $\mathcal{O}(\ell^2)$ terms in the above equation
\begin{align}
-\frac{\sigma_0}{f} \ddot{\tilde{z}} + \frac{\ell^2}{a^2} P_0 \tilde{z} =\mathcal{O}(\ell)\,,
\end{align}
so that $\omega^2 \sim -\ell^2 P_0/\sigma_0$ which is consistent with the scaling we obtained in the main text. The equation of motion for $w$ can be obtained by setting $\mu =\theta$ in Eq.~\eqref{eq:eom}, where the relevant terms have order $\omega^2 w,  \ell^2 w$ and $\partial h_{\mu\nu} \sim \mathcal{O}(\ell)$. As a result, we have $w \sim z \mathcal{O}(\ell^{-1})$.

\bibliography{ms}

\begin{thebibliography}{43}%
\makeatletter
\providecommand \@ifxundefined [1]{%
 \@ifx{#1\undefined}
}%
\providecommand \@ifnum [1]{%
 \ifnum #1\expandafter \@firstoftwo
 \else \expandafter \@secondoftwo
 \fi
}%
\providecommand \@ifx [1]{%
 \ifx #1\expandafter \@firstoftwo
 \else \expandafter \@secondoftwo
 \fi
}%
\providecommand \natexlab [1]{#1}%
\providecommand \enquote  [1]{``#1''}%
\providecommand \bibnamefont  [1]{#1}%
\providecommand \bibfnamefont [1]{#1}%
\providecommand \citenamefont [1]{#1}%
\providecommand \href@noop [0]{\@secondoftwo}%
\providecommand \href [0]{\begingroup \@sanitize@url \@href}%
\providecommand \@href[1]{\@@startlink{#1}\@@href}%
\providecommand \@@href[1]{\endgroup#1\@@endlink}%
\providecommand \@sanitize@url [0]{\catcode `\\12\catcode `\$12\catcode
  `\&12\catcode `\#12\catcode `\^12\catcode `\_12\catcode `\%12\relax}%
\providecommand \@@startlink[1]{}%
\providecommand \@@endlink[0]{}%
\providecommand \url  [0]{\begingroup\@sanitize@url \@url }%
\providecommand \@url [1]{\endgroup\@href {#1}{\urlprefix }}%
\providecommand \urlprefix  [0]{URL }%
\providecommand \Eprint [0]{\href }%
\providecommand \doibase [0]{http://dx.doi.org/}%
\providecommand \selectlanguage [0]{\@gobble}%
\providecommand \bibinfo  [0]{\@secondoftwo}%
\providecommand \bibfield  [0]{\@secondoftwo}%
\providecommand \translation [1]{[#1]}%
\providecommand \BibitemOpen [0]{}%
\providecommand \bibitemStop [0]{}%
\providecommand \bibitemNoStop [0]{.\EOS\space}%
\providecommand \EOS [0]{\spacefactor3000\relax}%
\providecommand \BibitemShut  [1]{\csname bibitem#1\endcsname}%
\let\auto@bib@innerbib\@empty
\bibitem [{\citenamefont {Abbott}\ \emph {et~al.}(2016)\citenamefont {Abbott}
  \emph {et~al.}}]{LIGOScientific:2016aoc}%
  \BibitemOpen
  \bibfield  {author} {\bibinfo {author} {\bibfnamefont {B.~P.}\ \bibnamefont
  {Abbott}} \emph {et~al.} (\bibinfo {collaboration} {LIGO Scientific,
  Virgo}),\ }\href {\doibase 10.1103/PhysRevLett.116.061102} {\bibfield
  {journal} {\bibinfo  {journal} {Phys. Rev. Lett.}\ }\textbf {\bibinfo
  {volume} {116}},\ \bibinfo {pages} {061102} (\bibinfo {year} {2016})},\
  \Eprint {http://arxiv.org/abs/1602.03837} {arXiv:1602.03837 [gr-qc]}
  \BibitemShut {NoStop}%
\bibitem [{\citenamefont {Abbott}\ \emph {et~al.}(2019)\citenamefont {Abbott}
  \emph {et~al.}}]{LIGOScientific:2018mvr}%
  \BibitemOpen
  \bibfield  {author} {\bibinfo {author} {\bibfnamefont {B.~P.}\ \bibnamefont
  {Abbott}} \emph {et~al.} (\bibinfo {collaboration} {LIGO Scientific,
  Virgo}),\ }\href {\doibase 10.1103/PhysRevX.9.031040} {\bibfield  {journal}
  {\bibinfo  {journal} {Phys. Rev. X}\ }\textbf {\bibinfo {volume} {9}},\
  \bibinfo {pages} {031040} (\bibinfo {year} {2019})},\ \Eprint
  {http://arxiv.org/abs/1811.12907} {arXiv:1811.12907 [astro-ph.HE]}
  \BibitemShut {NoStop}%
\bibitem [{\citenamefont {Abbott}\ \emph
  {et~al.}(2021{\natexlab{a}})\citenamefont {Abbott} \emph
  {et~al.}}]{LIGOScientific:2020ibl}%
  \BibitemOpen
  \bibfield  {author} {\bibinfo {author} {\bibfnamefont {R.}~\bibnamefont
  {Abbott}} \emph {et~al.} (\bibinfo {collaboration} {LIGO Scientific,
  Virgo}),\ }\href {\doibase 10.1103/PhysRevX.11.021053} {\bibfield  {journal}
  {\bibinfo  {journal} {Phys. Rev. X}\ }\textbf {\bibinfo {volume} {11}},\
  \bibinfo {pages} {021053} (\bibinfo {year} {2021}{\natexlab{a}})},\ \Eprint
  {http://arxiv.org/abs/2010.14527} {arXiv:2010.14527 [gr-qc]} \BibitemShut
  {NoStop}%
\bibitem [{\citenamefont {Abbott}\ \emph
  {et~al.}(2021{\natexlab{b}})\citenamefont {Abbott} \emph
  {et~al.}}]{LIGOScientific:2021djp}%
  \BibitemOpen
  \bibfield  {author} {\bibinfo {author} {\bibfnamefont {R.}~\bibnamefont
  {Abbott}} \emph {et~al.} (\bibinfo {collaboration} {LIGO Scientific, VIRGO,
  KAGRA}),\ }\href@noop {} {\  (\bibinfo {year} {2021}{\natexlab{b}})},\
  \Eprint {http://arxiv.org/abs/2111.03606} {arXiv:2111.03606 [gr-qc]}
  \BibitemShut {NoStop}%
\bibitem [{\citenamefont {{Event Horizon Telescope Collaboration}}\ \emph
  {et~al.}(2019)\citenamefont {{Event Horizon Telescope Collaboration}},
  \citenamefont {{Akiyama}}, \citenamefont {{Alberdi}}, \citenamefont {{Alef}},
  \citenamefont {{Asada}}, \citenamefont {{Azulay}}, \citenamefont {{Baczko}},
  \citenamefont {{Ball}}, \citenamefont {{Balokovi{\'c}}}, \citenamefont
  {{Barrett}},\ and\ \citenamefont {et~al.}}]{M87PaperI}%
  \BibitemOpen
  \bibfield  {author} {\bibinfo {author} {\bibnamefont {{Event Horizon
  Telescope Collaboration}}}, \bibinfo {author} {\bibfnamefont
  {K.}~\bibnamefont {{Akiyama}}}, \bibinfo {author} {\bibfnamefont
  {A.}~\bibnamefont {{Alberdi}}}, \bibinfo {author} {\bibfnamefont
  {W.}~\bibnamefont {{Alef}}}, \bibinfo {author} {\bibfnamefont
  {K.}~\bibnamefont {{Asada}}}, \bibinfo {author} {\bibfnamefont
  {R.}~\bibnamefont {{Azulay}}}, \bibinfo {author} {\bibfnamefont {A.-K.}\
  \bibnamefont {{Baczko}}}, \bibinfo {author} {\bibfnamefont {D.}~\bibnamefont
  {{Ball}}}, \bibinfo {author} {\bibfnamefont {M.}~\bibnamefont
  {{Balokovi{\'c}}}}, \bibinfo {author} {\bibfnamefont {J.}~\bibnamefont
  {{Barrett}}}, \ and\ \bibinfo {author} {\bibnamefont {et~al.}},\ }\href
  {\doibase 10.3847/2041-8213/ab0ec7} {\bibfield  {journal} {\bibinfo
  {journal} {The Astrophysical Journal Letters}\ }\textbf {\bibinfo {volume}
  {875}},\ \bibinfo {pages} {L1, (Paper I)} (\bibinfo {year}
  {2019})}\BibitemShut {NoStop}%
\bibitem [{\citenamefont {Akiyama}\ \emph {et~al.}(2022)\citenamefont {Akiyama}
  \emph {et~al.}}]{EventHorizonTelescope:2022xnr}%
  \BibitemOpen
  \bibfield  {author} {\bibinfo {author} {\bibfnamefont {K.}~\bibnamefont
  {Akiyama}} \emph {et~al.} (\bibinfo {collaboration} {Event Horizon
  Telescope}),\ }\href {\doibase 10.3847/2041-8213/ac6674} {\bibfield
  {journal} {\bibinfo  {journal} {Astrophys. J. Lett.}\ }\textbf {\bibinfo
  {volume} {930}},\ \bibinfo {pages} {L12} (\bibinfo {year}
  {2022})}\BibitemShut {NoStop}%
\bibitem [{\citenamefont {Sch{\"o}del}\ \emph {et~al.}(2002)\citenamefont
  {Sch{\"o}del}, \citenamefont {Ott}, \citenamefont {Genzel}, \citenamefont
  {Hofmann}, \citenamefont {Lehnert}, \citenamefont {Eckart}, \citenamefont
  {Mouawad}, \citenamefont {Alexander}, \citenamefont {Reid}, \citenamefont
  {Lenzen} \emph {et~al.}}]{schodel2002star}%
  \BibitemOpen
  \bibfield  {author} {\bibinfo {author} {\bibfnamefont {R.}~\bibnamefont
  {Sch{\"o}del}}, \bibinfo {author} {\bibfnamefont {T.}~\bibnamefont {Ott}},
  \bibinfo {author} {\bibfnamefont {R.}~\bibnamefont {Genzel}}, \bibinfo
  {author} {\bibfnamefont {R.}~\bibnamefont {Hofmann}}, \bibinfo {author}
  {\bibfnamefont {M.}~\bibnamefont {Lehnert}}, \bibinfo {author} {\bibfnamefont
  {A.}~\bibnamefont {Eckart}}, \bibinfo {author} {\bibfnamefont
  {N.}~\bibnamefont {Mouawad}}, \bibinfo {author} {\bibfnamefont
  {T.}~\bibnamefont {Alexander}}, \bibinfo {author} {\bibfnamefont
  {M.}~\bibnamefont {Reid}}, \bibinfo {author} {\bibfnamefont {R.}~\bibnamefont
  {Lenzen}},  \emph {et~al.},\ }\href@noop {} {\bibfield  {journal} {\bibinfo
  {journal} {Nature}\ }\textbf {\bibinfo {volume} {419}},\ \bibinfo {pages}
  {694} (\bibinfo {year} {2002})}\BibitemShut {NoStop}%
\bibitem [{\citenamefont {Berti}\ \emph {et~al.}(2016)\citenamefont {Berti},
  \citenamefont {Sesana}, \citenamefont {Barausse}, \citenamefont {Cardoso},\
  and\ \citenamefont {Belczynski}}]{Berti:2016lat}%
  \BibitemOpen
  \bibfield  {author} {\bibinfo {author} {\bibfnamefont {E.}~\bibnamefont
  {Berti}}, \bibinfo {author} {\bibfnamefont {A.}~\bibnamefont {Sesana}},
  \bibinfo {author} {\bibfnamefont {E.}~\bibnamefont {Barausse}}, \bibinfo
  {author} {\bibfnamefont {V.}~\bibnamefont {Cardoso}}, \ and\ \bibinfo
  {author} {\bibfnamefont {K.}~\bibnamefont {Belczynski}},\ }\href {\doibase
  10.1103/PhysRevLett.117.101102} {\bibfield  {journal} {\bibinfo  {journal}
  {Phys. Rev. Lett.}\ }\textbf {\bibinfo {volume} {117}},\ \bibinfo {pages}
  {101102} (\bibinfo {year} {2016})},\ \Eprint
  {http://arxiv.org/abs/1605.09286} {arXiv:1605.09286 [gr-qc]} \BibitemShut
  {NoStop}%
\bibitem [{\citenamefont {Yang}\ \emph {et~al.}(2017)\citenamefont {Yang},
  \citenamefont {Yagi}, \citenamefont {Blackman}, \citenamefont {Lehner},
  \citenamefont {Paschalidis}, \citenamefont {Pretorius},\ and\ \citenamefont
  {Yunes}}]{Yang:2017zxs}%
  \BibitemOpen
  \bibfield  {author} {\bibinfo {author} {\bibfnamefont {H.}~\bibnamefont
  {Yang}}, \bibinfo {author} {\bibfnamefont {K.}~\bibnamefont {Yagi}}, \bibinfo
  {author} {\bibfnamefont {J.}~\bibnamefont {Blackman}}, \bibinfo {author}
  {\bibfnamefont {L.}~\bibnamefont {Lehner}}, \bibinfo {author} {\bibfnamefont
  {V.}~\bibnamefont {Paschalidis}}, \bibinfo {author} {\bibfnamefont
  {F.}~\bibnamefont {Pretorius}}, \ and\ \bibinfo {author} {\bibfnamefont
  {N.}~\bibnamefont {Yunes}},\ }\href {\doibase 10.1103/PhysRevLett.118.161101}
  {\bibfield  {journal} {\bibinfo  {journal} {Phys. Rev. Lett.}\ }\textbf
  {\bibinfo {volume} {118}},\ \bibinfo {pages} {161101} (\bibinfo {year}
  {2017})},\ \Eprint {http://arxiv.org/abs/1701.05808} {arXiv:1701.05808
  [gr-qc]} \BibitemShut {NoStop}%
\bibitem [{\citenamefont {Punturo}\ \emph {et~al.}(2010)\citenamefont {Punturo}
  \emph {et~al.}}]{Punturo:2010zz}%
  \BibitemOpen
  \bibfield  {author} {\bibinfo {author} {\bibfnamefont {M.}~\bibnamefont
  {Punturo}} \emph {et~al.},\ }\href {\doibase 10.1088/0264-9381/27/19/194002}
  {\bibfield  {journal} {\bibinfo  {journal} {Class. Quant. Grav.}\ }\textbf
  {\bibinfo {volume} {27}},\ \bibinfo {pages} {194002} (\bibinfo {year}
  {2010})}\BibitemShut {NoStop}%
\bibitem [{\citenamefont {Reitze}\ \emph {et~al.}(2019)\citenamefont {Reitze},
  \citenamefont {Adhikari}, \citenamefont {Ballmer}, \citenamefont {Barish},
  \citenamefont {Barsotti}, \citenamefont {Billingsley}, \citenamefont {Brown},
  \citenamefont {Chen}, \citenamefont {Coyne}, \citenamefont {Eisenstein} \emph
  {et~al.}}]{reitze2019cosmic}%
  \BibitemOpen
  \bibfield  {author} {\bibinfo {author} {\bibfnamefont {D.}~\bibnamefont
  {Reitze}}, \bibinfo {author} {\bibfnamefont {R.~X.}\ \bibnamefont
  {Adhikari}}, \bibinfo {author} {\bibfnamefont {S.}~\bibnamefont {Ballmer}},
  \bibinfo {author} {\bibfnamefont {B.}~\bibnamefont {Barish}}, \bibinfo
  {author} {\bibfnamefont {L.}~\bibnamefont {Barsotti}}, \bibinfo {author}
  {\bibfnamefont {G.}~\bibnamefont {Billingsley}}, \bibinfo {author}
  {\bibfnamefont {D.~A.}\ \bibnamefont {Brown}}, \bibinfo {author}
  {\bibfnamefont {Y.}~\bibnamefont {Chen}}, \bibinfo {author} {\bibfnamefont
  {D.}~\bibnamefont {Coyne}}, \bibinfo {author} {\bibfnamefont
  {R.}~\bibnamefont {Eisenstein}},  \emph {et~al.},\ }\href@noop {} {\bibfield
  {journal} {\bibinfo  {journal} {arXiv preprint arXiv:1907.04833}\ } (\bibinfo
  {year} {2019})}\BibitemShut {NoStop}%
\bibitem [{\citenamefont {Amaro-Seoane}\ \emph {et~al.}(2017)\citenamefont
  {Amaro-Seoane} \emph {et~al.}}]{Audley:2017drz}%
  \BibitemOpen
  \bibfield  {author} {\bibinfo {author} {\bibfnamefont {P.}~\bibnamefont
  {Amaro-Seoane}} \emph {et~al.} (\bibinfo {collaboration} {LISA}),\
  }\href@noop {} {\  (\bibinfo {year} {2017})},\ \Eprint
  {http://arxiv.org/abs/1702.00786} {arXiv:1702.00786 [astro-ph.IM]}
  \BibitemShut {NoStop}%
\bibitem [{\citenamefont {Luo}\ \emph {et~al.}(2016)\citenamefont {Luo} \emph
  {et~al.}}]{TianQin:2015yph}%
  \BibitemOpen
  \bibfield  {author} {\bibinfo {author} {\bibfnamefont {J.}~\bibnamefont
  {Luo}} \emph {et~al.} (\bibinfo {collaboration} {TianQin}),\ }\href {\doibase
  10.1088/0264-9381/33/3/035010} {\bibfield  {journal} {\bibinfo  {journal}
  {Class. Quant. Grav.}\ }\textbf {\bibinfo {volume} {33}},\ \bibinfo {pages}
  {035010} (\bibinfo {year} {2016})},\ \Eprint
  {http://arxiv.org/abs/1512.02076} {arXiv:1512.02076 [astro-ph.IM]}
  \BibitemShut {NoStop}%
\bibitem [{\citenamefont {Cardoso}\ and\ \citenamefont
  {Pani}(2019)}]{Cardoso:2019rvt}%
  \BibitemOpen
  \bibfield  {author} {\bibinfo {author} {\bibfnamefont {V.}~\bibnamefont
  {Cardoso}}\ and\ \bibinfo {author} {\bibfnamefont {P.}~\bibnamefont {Pani}},\
  }\href {\doibase 10.1007/s41114-019-0020-4} {\bibfield  {journal} {\bibinfo
  {journal} {Living Rev. Rel.}\ }\textbf {\bibinfo {volume} {22}},\ \bibinfo
  {pages} {4} (\bibinfo {year} {2019})},\ \Eprint
  {http://arxiv.org/abs/1904.05363} {arXiv:1904.05363 [gr-qc]} \BibitemShut
  {NoStop}%
\bibitem [{\citenamefont {Haensel}\ and\ \citenamefont
  {Zdunik}(1989)}]{haensel1989submillisecond}%
  \BibitemOpen
  \bibfield  {author} {\bibinfo {author} {\bibfnamefont {P.}~\bibnamefont
  {Haensel}}\ and\ \bibinfo {author} {\bibfnamefont {J.}~\bibnamefont
  {Zdunik}},\ }\href@noop {} {\bibfield  {journal} {\bibinfo  {journal}
  {Nature}\ }\textbf {\bibinfo {volume} {340}},\ \bibinfo {pages} {617}
  (\bibinfo {year} {1989})}\BibitemShut {NoStop}%
\bibitem [{\citenamefont {Koranda}\ \emph {et~al.}(1997)\citenamefont
  {Koranda}, \citenamefont {Stergioulas},\ and\ \citenamefont
  {Friedman}}]{koranda1997upper}%
  \BibitemOpen
  \bibfield  {author} {\bibinfo {author} {\bibfnamefont {S.}~\bibnamefont
  {Koranda}}, \bibinfo {author} {\bibfnamefont {N.}~\bibnamefont
  {Stergioulas}}, \ and\ \bibinfo {author} {\bibfnamefont {J.~L.}\ \bibnamefont
  {Friedman}},\ }\href@noop {} {\bibfield  {journal} {\bibinfo  {journal} {The
  Astrophysical Journal}\ }\textbf {\bibinfo {volume} {488}},\ \bibinfo {pages}
  {799} (\bibinfo {year} {1997})}\BibitemShut {NoStop}%
\bibitem [{\citenamefont {Kesden}\ \emph {et~al.}(2005)\citenamefont {Kesden},
  \citenamefont {Gair},\ and\ \citenamefont {Kamionkowski}}]{Kesden:2004qx}%
  \BibitemOpen
  \bibfield  {author} {\bibinfo {author} {\bibfnamefont {M.}~\bibnamefont
  {Kesden}}, \bibinfo {author} {\bibfnamefont {J.}~\bibnamefont {Gair}}, \ and\
  \bibinfo {author} {\bibfnamefont {M.}~\bibnamefont {Kamionkowski}},\ }\href
  {\doibase 10.1103/PhysRevD.71.044015} {\bibfield  {journal} {\bibinfo
  {journal} {Phys. Rev. D}\ }\textbf {\bibinfo {volume} {71}},\ \bibinfo
  {pages} {044015} (\bibinfo {year} {2005})},\ \Eprint
  {http://arxiv.org/abs/astro-ph/0411478} {arXiv:astro-ph/0411478} \BibitemShut
  {NoStop}%
\bibitem [{\citenamefont {Bowers}\ and\ \citenamefont
  {Liang}(1974)}]{Bowers:1974tgi}%
  \BibitemOpen
  \bibfield  {author} {\bibinfo {author} {\bibfnamefont {R.~L.}\ \bibnamefont
  {Bowers}}\ and\ \bibinfo {author} {\bibfnamefont {E.~P.~T.}\ \bibnamefont
  {Liang}},\ }\href {\doibase 10.1086/152760} {\bibfield  {journal} {\bibinfo
  {journal} {Astrophys. J.}\ }\textbf {\bibinfo {volume} {188}},\ \bibinfo
  {pages} {657} (\bibinfo {year} {1974})}\BibitemShut {NoStop}%
\bibitem [{\citenamefont {Bayin}(1982)}]{Bayin:1982vw}%
  \BibitemOpen
  \bibfield  {author} {\bibinfo {author} {\bibfnamefont {S.~S.}\ \bibnamefont
  {Bayin}},\ }\href {\doibase 10.1103/PhysRevD.26.1262} {\bibfield  {journal}
  {\bibinfo  {journal} {Phys. Rev. D}\ }\textbf {\bibinfo {volume} {26}},\
  \bibinfo {pages} {1262} (\bibinfo {year} {1982})}\BibitemShut {NoStop}%
\bibitem [{\citenamefont {Dev}\ and\ \citenamefont
  {Gleiser}(2002)}]{Dev:2000gt}%
  \BibitemOpen
  \bibfield  {author} {\bibinfo {author} {\bibfnamefont {K.}~\bibnamefont
  {Dev}}\ and\ \bibinfo {author} {\bibfnamefont {M.}~\bibnamefont {Gleiser}},\
  }\href {\doibase 10.1023/A:1020707906543} {\bibfield  {journal} {\bibinfo
  {journal} {Gen. Rel. Grav.}\ }\textbf {\bibinfo {volume} {34}},\ \bibinfo
  {pages} {1793} (\bibinfo {year} {2002})},\ \Eprint
  {http://arxiv.org/abs/astro-ph/0012265} {arXiv:astro-ph/0012265} \BibitemShut
  {NoStop}%
\bibitem [{\citenamefont {Mak}\ and\ \citenamefont {Harko}(2003)}]{Mak:2001eb}%
  \BibitemOpen
  \bibfield  {author} {\bibinfo {author} {\bibfnamefont {M.~K.}\ \bibnamefont
  {Mak}}\ and\ \bibinfo {author} {\bibfnamefont {T.}~\bibnamefont {Harko}},\
  }\href {\doibase 10.1098/rspa.2002.1014} {\bibfield  {journal} {\bibinfo
  {journal} {Proc. Roy. Soc. Lond. A}\ }\textbf {\bibinfo {volume} {459}},\
  \bibinfo {pages} {393} (\bibinfo {year} {2003})},\ \Eprint
  {http://arxiv.org/abs/gr-qc/0110103} {arXiv:gr-qc/0110103} \BibitemShut
  {NoStop}%
\bibitem [{\citenamefont {Herrera}\ \emph {et~al.}(2004)\citenamefont
  {Herrera}, \citenamefont {Di~Prisco}, \citenamefont {Martin}, \citenamefont
  {Ospino}, \citenamefont {Santos},\ and\ \citenamefont
  {Troconis}}]{Herrera:2004xc}%
  \BibitemOpen
  \bibfield  {author} {\bibinfo {author} {\bibfnamefont {L.}~\bibnamefont
  {Herrera}}, \bibinfo {author} {\bibfnamefont {A.}~\bibnamefont {Di~Prisco}},
  \bibinfo {author} {\bibfnamefont {J.}~\bibnamefont {Martin}}, \bibinfo
  {author} {\bibfnamefont {J.}~\bibnamefont {Ospino}}, \bibinfo {author}
  {\bibfnamefont {N.~O.}\ \bibnamefont {Santos}}, \ and\ \bibinfo {author}
  {\bibfnamefont {O.}~\bibnamefont {Troconis}},\ }\href {\doibase
  10.1103/PhysRevD.69.084026} {\bibfield  {journal} {\bibinfo  {journal} {Phys.
  Rev. D}\ }\textbf {\bibinfo {volume} {69}},\ \bibinfo {pages} {084026}
  (\bibinfo {year} {2004})},\ \Eprint {http://arxiv.org/abs/gr-qc/0403006}
  {arXiv:gr-qc/0403006} \BibitemShut {NoStop}%
\bibitem [{\citenamefont {Yagi}\ and\ \citenamefont
  {Yunes}(2015)}]{Yagi:2015hda}%
  \BibitemOpen
  \bibfield  {author} {\bibinfo {author} {\bibfnamefont {K.}~\bibnamefont
  {Yagi}}\ and\ \bibinfo {author} {\bibfnamefont {N.}~\bibnamefont {Yunes}},\
  }\href {\doibase 10.1103/PhysRevD.91.123008} {\bibfield  {journal} {\bibinfo
  {journal} {Phys. Rev. D}\ }\textbf {\bibinfo {volume} {91}},\ \bibinfo
  {pages} {123008} (\bibinfo {year} {2015})},\ \Eprint
  {http://arxiv.org/abs/1503.02726} {arXiv:1503.02726 [gr-qc]} \BibitemShut
  {NoStop}%
\bibitem [{\citenamefont {Raposo}\ \emph {et~al.}(2019)\citenamefont {Raposo},
  \citenamefont {Pani}, \citenamefont {Bezares}, \citenamefont {Palenzuela},\
  and\ \citenamefont {Cardoso}}]{Raposo:2018rjn}%
  \BibitemOpen
  \bibfield  {author} {\bibinfo {author} {\bibfnamefont {G.}~\bibnamefont
  {Raposo}}, \bibinfo {author} {\bibfnamefont {P.}~\bibnamefont {Pani}},
  \bibinfo {author} {\bibfnamefont {M.}~\bibnamefont {Bezares}}, \bibinfo
  {author} {\bibfnamefont {C.}~\bibnamefont {Palenzuela}}, \ and\ \bibinfo
  {author} {\bibfnamefont {V.}~\bibnamefont {Cardoso}},\ }\href {\doibase
  10.1103/PhysRevD.99.104072} {\bibfield  {journal} {\bibinfo  {journal} {Phys.
  Rev. D}\ }\textbf {\bibinfo {volume} {99}},\ \bibinfo {pages} {104072}
  (\bibinfo {year} {2019})},\ \Eprint {http://arxiv.org/abs/1811.07917}
  {arXiv:1811.07917 [gr-qc]} \BibitemShut {NoStop}%
\bibitem [{\citenamefont {Alho}\ \emph {et~al.}(2022)\citenamefont {Alho},
  \citenamefont {Nat\'ario}, \citenamefont {Pani},\ and\ \citenamefont
  {Raposo}}]{Alho:2022bki}%
  \BibitemOpen
  \bibfield  {author} {\bibinfo {author} {\bibfnamefont {A.}~\bibnamefont
  {Alho}}, \bibinfo {author} {\bibfnamefont {J.}~\bibnamefont {Nat\'ario}},
  \bibinfo {author} {\bibfnamefont {P.}~\bibnamefont {Pani}}, \ and\ \bibinfo
  {author} {\bibfnamefont {G.}~\bibnamefont {Raposo}},\ }\href@noop {} {\
  (\bibinfo {year} {2022})},\ \Eprint {http://arxiv.org/abs/2202.00043}
  {arXiv:2202.00043 [gr-qc]} \BibitemShut {NoStop}%
\bibitem [{\citenamefont {{Mazur}}\ and\ \citenamefont
  {{Mottola}}(2001)}]{Mazur2001}%
  \BibitemOpen
  \bibfield  {author} {\bibinfo {author} {\bibfnamefont {P.~O.}\ \bibnamefont
  {{Mazur}}}\ and\ \bibinfo {author} {\bibfnamefont {E.}~\bibnamefont
  {{Mottola}}},\ }\href@noop {} {\bibfield  {journal} {\bibinfo  {journal}
  {arXiv e-prints}\ ,\ \bibinfo {eid} {gr-qc/0109035}} (\bibinfo {year}
  {2001})},\ \Eprint {http://arxiv.org/abs/gr-qc/0109035} {arXiv:gr-qc/0109035
  [gr-qc]} \BibitemShut {NoStop}%
\bibitem [{\citenamefont {{Visser}}(1989{\natexlab{a}})}]{Visser1989}%
  \BibitemOpen
  \bibfield  {author} {\bibinfo {author} {\bibfnamefont {M.}~\bibnamefont
  {{Visser}}},\ }\href {\doibase 10.1103/PhysRevD.39.3182} {\bibfield
  {journal} {\bibinfo  {journal} {\prd}\ }\textbf {\bibinfo {volume} {39}},\
  \bibinfo {pages} {3182} (\bibinfo {year} {1989}{\natexlab{a}})},\ \Eprint
  {http://arxiv.org/abs/0809.0907} {arXiv:0809.0907 [gr-qc]} \BibitemShut
  {NoStop}%
\bibitem [{\citenamefont {{Visser}}(1989{\natexlab{b}})}]{Visser1989b}%
  \BibitemOpen
  \bibfield  {author} {\bibinfo {author} {\bibfnamefont {M.}~\bibnamefont
  {{Visser}}},\ }\href {\doibase 10.1016/0550-3213(89)90100-4} {\bibfield
  {journal} {\bibinfo  {journal} {Nuclear Physics B}\ }\textbf {\bibinfo
  {volume} {328}},\ \bibinfo {pages} {203} (\bibinfo {year}
  {1989}{\natexlab{b}})},\ \Eprint {http://arxiv.org/abs/0809.0927}
  {arXiv:0809.0927 [gr-qc]} \BibitemShut {NoStop}%
\bibitem [{\citenamefont {Holdom}\ and\ \citenamefont
  {Ren}(2017)}]{Holdom:2016nek}%
  \BibitemOpen
  \bibfield  {author} {\bibinfo {author} {\bibfnamefont {B.}~\bibnamefont
  {Holdom}}\ and\ \bibinfo {author} {\bibfnamefont {J.}~\bibnamefont {Ren}},\
  }\href {\doibase 10.1103/PhysRevD.95.084034} {\bibfield  {journal} {\bibinfo
  {journal} {Phys. Rev. D}\ }\textbf {\bibinfo {volume} {95}},\ \bibinfo
  {pages} {084034} (\bibinfo {year} {2017})},\ \Eprint
  {http://arxiv.org/abs/1612.04889} {arXiv:1612.04889 [gr-qc]} \BibitemShut
  {NoStop}%
\bibitem [{\citenamefont {Kaplan}\ and\ \citenamefont
  {Rajendran}(2019)}]{Kaplan:2018dqx}%
  \BibitemOpen
  \bibfield  {author} {\bibinfo {author} {\bibfnamefont {D.~E.}\ \bibnamefont
  {Kaplan}}\ and\ \bibinfo {author} {\bibfnamefont {S.}~\bibnamefont
  {Rajendran}},\ }\href {\doibase 10.1103/PhysRevD.99.044033} {\bibfield
  {journal} {\bibinfo  {journal} {Phys. Rev. D}\ }\textbf {\bibinfo {volume}
  {99}},\ \bibinfo {pages} {044033} (\bibinfo {year} {2019})},\ \Eprint
  {http://arxiv.org/abs/1812.00536} {arXiv:1812.00536 [hep-th]} \BibitemShut
  {NoStop}%
\bibitem [{\citenamefont {Gregorian}\ \emph {et~al.}(2015)\citenamefont
  {Gregorian} \emph {et~al.}}]{gregorian2015nonradial}%
  \BibitemOpen
  \bibfield  {author} {\bibinfo {author} {\bibfnamefont {P.}~\bibnamefont
  {Gregorian}} \emph {et~al.},\ }\emph {\bibinfo {title} {Nonradial neutron
  star oscillations}},\ \href@noop {} {Master's thesis} (\bibinfo {year}
  {2015})\BibitemShut {NoStop}%
\bibitem [{\citenamefont {Poisson}\ and\ \citenamefont
  {Will}(2014)}]{poisson2014gravity}%
  \BibitemOpen
  \bibfield  {author} {\bibinfo {author} {\bibfnamefont {E.}~\bibnamefont
  {Poisson}}\ and\ \bibinfo {author} {\bibfnamefont {C.~M.}\ \bibnamefont
  {Will}},\ }\href@noop {} {\bibfield  {journal} {\bibinfo  {journal}
  {Gravity}\ } (\bibinfo {year} {2014})}\BibitemShut {NoStop}%
\bibitem [{\citenamefont {Poisson}\ and\ \citenamefont
  {Visser}(1995)}]{Poisson:1995sv}%
  \BibitemOpen
  \bibfield  {author} {\bibinfo {author} {\bibfnamefont {E.}~\bibnamefont
  {Poisson}}\ and\ \bibinfo {author} {\bibfnamefont {M.}~\bibnamefont
  {Visser}},\ }\href {\doibase 10.1103/PhysRevD.52.7318} {\bibfield  {journal}
  {\bibinfo  {journal} {Phys. Rev. D}\ }\textbf {\bibinfo {volume} {52}},\
  \bibinfo {pages} {7318} (\bibinfo {year} {1995})},\ \Eprint
  {http://arxiv.org/abs/gr-qc/9506083} {arXiv:gr-qc/9506083} \BibitemShut
  {NoStop}%
\bibitem [{\citenamefont {Barack}\ and\ \citenamefont
  {Lousto}(2005)}]{Barack:2005nr}%
  \BibitemOpen
  \bibfield  {author} {\bibinfo {author} {\bibfnamefont {L.}~\bibnamefont
  {Barack}}\ and\ \bibinfo {author} {\bibfnamefont {C.~O.}\ \bibnamefont
  {Lousto}},\ }\href {\doibase 10.1103/PhysRevD.72.104026} {\bibfield
  {journal} {\bibinfo  {journal} {Phys. Rev. D}\ }\textbf {\bibinfo {volume}
  {72}},\ \bibinfo {pages} {104026} (\bibinfo {year} {2005})},\ \Eprint
  {http://arxiv.org/abs/gr-qc/0510019} {arXiv:gr-qc/0510019} \BibitemShut
  {NoStop}%
\bibitem [{\citenamefont {Danielsson}\ \emph {et~al.}(2017)\citenamefont
  {Danielsson}, \citenamefont {Dibitetto},\ and\ \citenamefont
  {Giri}}]{Danielsson:2017riq}%
  \BibitemOpen
  \bibfield  {author} {\bibinfo {author} {\bibfnamefont {U.~H.}\ \bibnamefont
  {Danielsson}}, \bibinfo {author} {\bibfnamefont {G.}~\bibnamefont
  {Dibitetto}}, \ and\ \bibinfo {author} {\bibfnamefont {S.}~\bibnamefont
  {Giri}},\ }\href {\doibase 10.1007/JHEP10(2017)171} {\bibfield  {journal}
  {\bibinfo  {journal} {JHEP}\ }\textbf {\bibinfo {volume} {10}},\ \bibinfo
  {pages} {171} (\bibinfo {year} {2017})},\ \Eprint
  {http://arxiv.org/abs/1705.10172} {arXiv:1705.10172 [hep-th]} \BibitemShut
  {NoStop}%
\bibitem [{\citenamefont {Danielsson}\ \emph {et~al.}(2021)\citenamefont
  {Danielsson}, \citenamefont {Lehner},\ and\ \citenamefont
  {Pretorius}}]{Danielsson:2021ykm}%
  \BibitemOpen
  \bibfield  {author} {\bibinfo {author} {\bibfnamefont {U.}~\bibnamefont
  {Danielsson}}, \bibinfo {author} {\bibfnamefont {L.}~\bibnamefont {Lehner}},
  \ and\ \bibinfo {author} {\bibfnamefont {F.}~\bibnamefont {Pretorius}},\
  }\href {\doibase 10.1103/PhysRevD.104.124011} {\bibfield  {journal} {\bibinfo
   {journal} {Phys. Rev. D}\ }\textbf {\bibinfo {volume} {104}},\ \bibinfo
  {pages} {124011} (\bibinfo {year} {2021})},\ \Eprint
  {http://arxiv.org/abs/2109.09814} {arXiv:2109.09814 [gr-qc]} \BibitemShut
  {NoStop}%
\bibitem [{\citenamefont {Pani}\ \emph {et~al.}(2009)\citenamefont {Pani},
  \citenamefont {Berti}, \citenamefont {Cardoso}, \citenamefont {Chen},\ and\
  \citenamefont {Norte}}]{Pani:2009ss}%
  \BibitemOpen
  \bibfield  {author} {\bibinfo {author} {\bibfnamefont {P.}~\bibnamefont
  {Pani}}, \bibinfo {author} {\bibfnamefont {E.}~\bibnamefont {Berti}},
  \bibinfo {author} {\bibfnamefont {V.}~\bibnamefont {Cardoso}}, \bibinfo
  {author} {\bibfnamefont {Y.}~\bibnamefont {Chen}}, \ and\ \bibinfo {author}
  {\bibfnamefont {R.}~\bibnamefont {Norte}},\ }\href {\doibase
  10.1103/PhysRevD.80.124047} {\bibfield  {journal} {\bibinfo  {journal} {Phys.
  Rev. D}\ }\textbf {\bibinfo {volume} {80}},\ \bibinfo {pages} {124047}
  (\bibinfo {year} {2009})},\ \Eprint {http://arxiv.org/abs/0909.0287}
  {arXiv:0909.0287 [gr-qc]} \BibitemShut {NoStop}%
\bibitem [{\citenamefont {Cardoso}\ \emph
  {et~al.}(2016{\natexlab{a}})\citenamefont {Cardoso}, \citenamefont
  {Franzin},\ and\ \citenamefont {Pani}}]{Cardoso:2016rao}%
  \BibitemOpen
  \bibfield  {author} {\bibinfo {author} {\bibfnamefont {V.}~\bibnamefont
  {Cardoso}}, \bibinfo {author} {\bibfnamefont {E.}~\bibnamefont {Franzin}}, \
  and\ \bibinfo {author} {\bibfnamefont {P.}~\bibnamefont {Pani}},\ }\href
  {\doibase 10.1103/PhysRevLett.116.171101} {\bibfield  {journal} {\bibinfo
  {journal} {Phys. Rev. Lett.}\ }\textbf {\bibinfo {volume} {116}},\ \bibinfo
  {pages} {171101} (\bibinfo {year} {2016}{\natexlab{a}})},\ \bibinfo {note}
  {[Erratum: Phys.Rev.Lett. 117, 089902 (2016)]},\ \Eprint
  {http://arxiv.org/abs/1602.07309} {arXiv:1602.07309 [gr-qc]} \BibitemShut
  {NoStop}%
\bibitem [{\citenamefont {Cardoso}\ \emph
  {et~al.}(2016{\natexlab{b}})\citenamefont {Cardoso}, \citenamefont {Hopper},
  \citenamefont {Macedo}, \citenamefont {Palenzuela},\ and\ \citenamefont
  {Pani}}]{Cardoso:2016oxy}%
  \BibitemOpen
  \bibfield  {author} {\bibinfo {author} {\bibfnamefont {V.}~\bibnamefont
  {Cardoso}}, \bibinfo {author} {\bibfnamefont {S.}~\bibnamefont {Hopper}},
  \bibinfo {author} {\bibfnamefont {C.~F.~B.}\ \bibnamefont {Macedo}}, \bibinfo
  {author} {\bibfnamefont {C.}~\bibnamefont {Palenzuela}}, \ and\ \bibinfo
  {author} {\bibfnamefont {P.}~\bibnamefont {Pani}},\ }\href {\doibase
  10.1103/PhysRevD.94.084031} {\bibfield  {journal} {\bibinfo  {journal} {Phys.
  Rev. D}\ }\textbf {\bibinfo {volume} {94}},\ \bibinfo {pages} {084031}
  (\bibinfo {year} {2016}{\natexlab{b}})},\ \Eprint
  {http://arxiv.org/abs/1608.08637} {arXiv:1608.08637 [gr-qc]} \BibitemShut
  {NoStop}%
\bibitem [{\citenamefont {Bonga}\ and\ \citenamefont
  {Yang}(2021)}]{Bonga:2021ouq}%
  \BibitemOpen
  \bibfield  {author} {\bibinfo {author} {\bibfnamefont {B.}~\bibnamefont
  {Bonga}}\ and\ \bibinfo {author} {\bibfnamefont {H.}~\bibnamefont {Yang}},\
  }\href {\doibase 10.1103/PhysRevD.104.084040} {\bibfield  {journal} {\bibinfo
   {journal} {Phys. Rev. D}\ }\textbf {\bibinfo {volume} {104}},\ \bibinfo
  {pages} {084040} (\bibinfo {year} {2021})},\ \Eprint
  {http://arxiv.org/abs/2106.08342} {arXiv:2106.08342 [gr-qc]} \BibitemShut
  {NoStop}%
\bibitem [{Note1()}]{Note1}%
  \BibitemOpen
  \bibinfo {note} {The purpose of \cite {Bonga:2021ouq} is to theoretically
  demonstrate that a compact object can have the same set of multipole moments
  of Kerr, instead of proposing its astrophysical relevance.}\BibitemShut
  {Stop}%
\bibitem [{Note2()}]{Note2}%
  \BibitemOpen
  \bibinfo {note} {Private communications with Eric Poisson.}\BibitemShut
  {Stop}%
\bibitem [{\citenamefont {Pani}(2015)}]{Pani:2015tga}%
  \BibitemOpen
  \bibfield  {author} {\bibinfo {author} {\bibfnamefont {P.}~\bibnamefont
  {Pani}},\ }\href {\doibase 10.1103/PhysRevD.95.049902} {\bibfield  {journal}
  {\bibinfo  {journal} {Phys. Rev. D}\ }\textbf {\bibinfo {volume} {92}},\
  \bibinfo {pages} {124030} (\bibinfo {year} {2015})},\ \bibinfo {note}
  {[Erratum: Phys.Rev.D 95, 049902 (2017)]},\ \Eprint
  {http://arxiv.org/abs/1506.06050} {arXiv:1506.06050 [gr-qc]} \BibitemShut
  {NoStop}%
\end{thebibliography}%

\end{document}